\documentclass[reprint,aps,prb,amsmath,amssymb]{revtex4-1}

\usepackage[]{graphicx}
\usepackage[]{units}
\usepackage{times}
\usepackage{bm}
\usepackage[ulem=normalem]{changes}
\usepackage[normalem]{ulem}		
\usepackage{color}
\usepackage{xr} 				
\usepackage{cancel}

\usepackage[utf8]{inputenc}
\usepackage[T1]{fontenc}
\definechangesauthor[color=blue]{SM}

\newcommand{\re}{\Re\textnormal{e}}
\newcommand{\im}{\Im\textnormal{m}}

\newcommand{\blue}[1]{{\color[rgb]{0,0,1}{#1}}} 

\newcommand{\kp}{\mathbf{k}_{\parallel}}
\newcommand{\qp}{\mathbf{Q}_{\parallel}}
\newcommand{\gp}{\mathbf{G}_{\parallel}}

\renewcommand{\emph}{\textit}

\newcommand{\one}{1\!\!1}

\begin{document}

\title{Theory of X-ray absorption spectroscopy: a microscopic Bloch equation approach for two-dimensional solid states}
\author{Dominik Christiansen$^{1}$}
\author{Malte Selig$^{1}$}
\author{Jens Biegert$^{2,3,4}$}
\author{Andreas Knorr$^{1}$}
\affiliation{$^{1}$Institut f\"ur Theoretische Physik, Nichtlineare Optik und Quantenelektronik, Technische Universit\"at Berlin, 10623 Berlin, Germany}
\affiliation{$^{2}$ICFO-Institut de Ciencies Fotoniques, The Barcelona Institute of Science and Technology, 08860 Castelldefels (Barcelona), Spain}
\affiliation{$^{3}$ICREA-Instituci\'{o} Catalana de Recerca i Estudis Avan\c{c}ats, Barcelona, Spain}
\affiliation{$^{4}$Fritz-Haber Institute of the Max Planck Society, Berlin, Germany}

\begin{abstract}
We develop a self-consistent Maxwell-Bloch formalism for the interaction of X-rays with two-dimensional crystalline materials by incorporating the Bloch theorem and Coulomb many-body interaction. This formalism is illustrated for graphene, by calculating the polarization-dependent XANES, formulating expressions for the radiative and Meinter-Auger recombination of core-holes, and the discussion of microscopic insights into the spectral oscillations of EXAFS beyond point scattering theory. In particular, the correct inclusion of lattice periodicity in our evaluation allows us to assign so far uninterpreted spectral features in the Fourier transformed EXAFS spectrum.
\end{abstract}

\maketitle


\section{Introduction}
At the beginning of the last century X-ray absorption spectroscopy (XAS) has been the key instrument to acquire knowledge about atomic energy levels \cite{Bohr1913,Sommerfeld1916,Moseley1913,Lundquist1925,Coster1927} and later for the discovery and systematization of rare-earth elements \cite{Roehler1987}.
The constant development of intense synchrotron radiation sources during these times led to a steady growth in X-ray experiments \cite{Lytle1965,Eisenberger1978,Bordwehr1989,Sauer2008,Young2018,Gu2021}. At the end of the century a connection between the X-ray absorption spectrum and the local structure of the investigated material has been drawn. This promoted XAS from a spectroscopic to a structural characterization technique \cite{Sayers1970,Sayers1971}: X-ray absorption spectroscopy, probing the core electronic states of atomic systems, is a highly element-sensitive and environment-specific spectroscopic technique with applications in atom and molecular physics, chemistry, biology and material science \cite{Eisenberger1978_2,Yano2009,Kraus2018,Diller2017,Hua2019}. For this, a large range of tunable X-ray synchrotron sources \cite{Esarey1993,Schoenlein2000,Buades2018} sensitive for most elements in the periodic table are available.

In this contribution, we develop, based on a Bloch function approach in the Heisenberg equation formalism, a microscopic description of X-ray induced electronic transitions and ionization for solid state two-dimensional semiconductors, Dirac semimetals or insulators \cite{Novoselov2005,Balleste2011,Xu2013,Gu2018}. We include nonlinear excitation effects, discuss X-ray absorption dynamics in solids beyond Fermi's golden rule, and include explicitly the lattice periodicity of crystalline solids. In particular, we shine light on the microscopic origin of the spectral oscillations observed in absorption and identify quantum-interference peaks, which can not be described in point scattering theory \cite{Buades2018}.
\begin{figure}[t]
\begin{center}
\includegraphics[width=\linewidth]{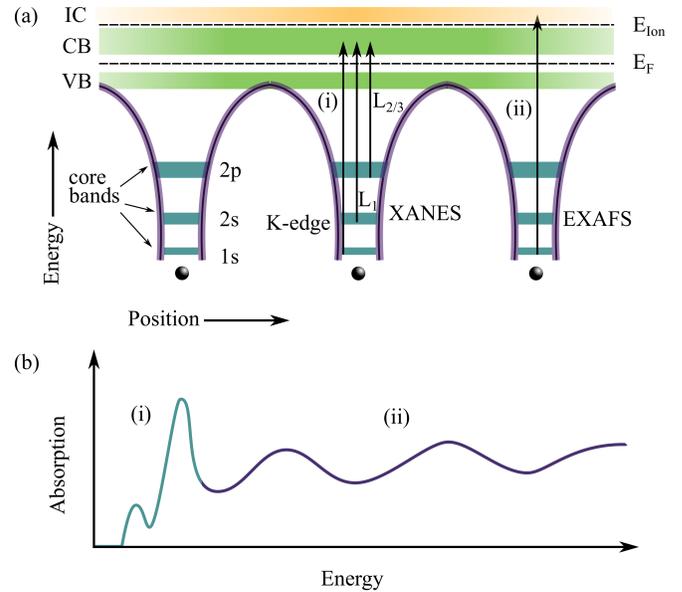}
\end{center}
\caption{(a) Lattice periodic potential resulting from periodically arranged atoms in a solid. Sketched are exemplary flat core states, continuous bands forming higher lying valence bands (VB) and conduction bands (CB) and the ionization continuum (IC) above the ionization threshold. Transitions into the conduction bands and the ionization continuum form the XANES and EXAFS, respectively. (b) Sketch of a typical XAS spectrum with its (i) XANES and (ii) EXAFS part.}
\label{fig:sketch}
\end{figure}

Typical experimental X-ray absorption spectra exhibit two general features: (i) resonance lines in the absorption coefficient at specific energies below the ionization threshold and (ii) a continuous absorption coefficient above the ionization threshold over a large photon energy range \citep{Lindsay1928,Coster1929,Kievit1930,Norman1986,Rehr2000,Buades2018}, including spectral oscillations. The latter appear only for molecules or crystals but not for single atoms. Figure \ref{fig:sketch}(a) shows the possible excitations, while Fig. \ref{fig:sketch}(b) sketches the corresponding spectral features.

The first feature (i) is related to the transition energy of most-inner core shell electrons to unoccupied states in the conduction band (CB) or partly filled valence bands (VB). This characteristic resonant feature is usually referred to as absorption edge and the spectroscopic technique exploiting such transitions is called X-ray absorption near-edge structure (XANES) or near-edge X-ray absorption fine structure (NEXAFS). In the existing literature, to describe XANES, typically Fermi's golden rule for the transition probability from core bands to unoccupied bands under illumination with a X-ray frequency $\omega$ is used \cite{Stohr1992,Chowdhury2012,Schnohr2015}:
\begin{align}
\alpha(\omega)=\frac{\omega\pi}{\epsilon_0cn}\int\int d\mathbf{k}~ d\mathbf{k}' ~ |\mathbf{e}\cdot\mathbf{d}_{\mathbf{k},\mathbf{k}'}|^2 \delta(E_{\mathbf{k}'}-E_{\mathbf{k}}-\hbar\omega) \;. \label{eq:fermi}
\end{align}
Here, the transition probability $\alpha(\omega)$ involving initial core states $\mathbf{k}$ to final conduction band states $\mathbf{k}'$ is determined by the product of the transition dipole moment $\mathbf{d}_{\mathbf{k},\mathbf{k}'}$ projected on the incident X-ray polarization $\mathbf{e}$ and an energy conserving delta-function during the transition. The prefactor incorporates the dielectric constant $\epsilon_0$, speed of light in vacuum $c$, and the refractive index $n$ of the material.

The part of the absorption spectrum containing spectral oscillations above the ionization energy (ii) is named extended X-ray absorption fine structure (EXAFS) spectrum. It results from the transition of core electrons to the ionization continuum (IC) above the ionization threshold of the material, cf. Fig. \ref{fig:sketch}(a). The explanation of the EXAFS spectrum is based on the theory introduced by Kronig for molecular gases\cite{Kronig1931,Kronig1932}. This description explains the oscillations in the spectrum as interference effects emerging from secondary photoelectron waves emitted by an X-ray absorbing atom and waves back scattered by neighboring atoms. The corresponding parametrization of the absorption by Sayers \textit{et al.} \cite{Sayers1970,Sayers1971}
\begin{align}
\alpha_{k}&=S_0^2\sum_i N_i \frac{|f_{i}(k)|}{kR_i^2} \sin\left(2kR_i+\phi_i\right) e^{-\frac{2R_i}{\lambda_k}}e^{-2\sigma_i^2k^2} \label{eq:scat}
\end{align}
has become standard in describing EXAFS. Equation \eqref{eq:scat} describes the absorption coefficient as function of the electronic wave number $k$ of the ionized, free electron. Here, for X-ray energies higher than the ionization energy $E_{Ion}$ of the material (EXAFS), the excess energy is transfered to the excited photoelectron in form of kinetic energy. The wave number of the emitted electrons is related to the X-ray excitation frequency $\omega$ by $k^2=2m_0(\hbar\omega-E_{Ion})/\hbar^2$ with the free (vacuum) electron mass $m_0$. Here, $\lambda_k$ stands for the X-ray wavelength. The structural parameters are the number of equivalent scatterers $N_i$ of type $i$, the interatomic distances $R_i$, and the bond length $\sigma_i$. The Debye-Waller factor accounts for thermally or disorder induced changes in the bond length. The quantity $|f_i(k)|$ describes the back scattering amplitude, resulting from back scattering of neighboring atoms, and $S_0$ describes the reduction factor due to multielectron processes. The exponential factor $\exp(-2R_i/\lambda_k)$ accounts for the finite lifetime of the photoelectron, which is only scattered elastically over a short distance. Finally, $\phi_i$ stand for phase shifts due to the initially excited and the back scattering atom \cite{Sayers1971,Rehr2000}. The wave functions used to obtain this formula are not lattice periodic and therefore do not exploit the Bloch theorem in crystalline matter. While this approach might be appropriate for molecules, where only a few $R_i$ have to be accounted for, a X-ray absorption theory for solids should take the lattice periodicity into account. 

In this manuscript, to address recent advances in pulsed X-ray spectroscopy\cite{Popmintchev2012,Teichmann2016,Pertot2017} of solids we develop a dynamical approach to account for the description of ultrafast many-body solid state phenomena in time resolved X-ray spectroscopy \cite{Angela2015,Picon2019,Chang2021}: We develop a microscopic, temporal resolved formulation of the underlying mechanisms of the X-ray absorption in a two-dimensional solid. For this, we use the method of second quantization to also have a basis to sequentially include many-body interaction at a later stage of theory. In particular, we shine light on the origin of peaks in the Fourier transformed EXAFS spectrum, which are not captured by the point scattering theory but observable in experiment \cite{Buades2018}. Incorporating Maxwell equations for the X-ray field a self-consistent coupling between the light field and the microscopic core-transition including radiative damping and lineshifts can be achieved. Including also many-body interactions the resulting Maxwell-Bloch formalism  provides a powerful tool to investigate core excitations induced by classical light in solids.

The manuscript is organized as follows: We start in Section \ref{sec:Hamilton} by setting up the Hamiltonian on the basis of single-particle electronic wave functions and their interaction matrix elements for the XANES and EXAFS processes. This section is separated into two parts: the first is using a general Bloch wave approach suited for periodic atomically-thin solids, while the second part bridges the Bloch description with a tight binding approximation often used in electronic structure theory. In Section \ref{sec:Bloch} we derive the dynamical Bloch equations for the X-ray induced electronic transitions. Here, we do not only consider the X-ray induced core excitations but also display the Coulomb-induced mean-field Hartree-Fock and relaxation channels of the excited core electrons. In particular, the equations of motion include the spatial resolution of the X-ray radiation. In Section \ref{sec:Absorption}, we solve the wave equation to couple self-consistently the Maxwell equations to the non-local microscopic excitations, including nonlinear effects and finally apply in Section \ref{sec:Graphene} the developed theory to the exemplary material of graphene.

\section{Solid state electron--X-ray interaction} \label{sec:Hamilton}
\subsection{Hamiltonian} \label{sec:Hamilton1}
To develop a microscopic theory of electron-X-ray interaction in solids, we start by deriving the interaction Hamiltonian. A schematic picture of the electronic structure is sketched in Fig. \ref{fig:sketch}(a). Typically, electrons in crystalline solids can be separated into two groups: core and valence electrons. Core electrons occupy filled orbitals and are spatially localized at the nuclei, both contributing to core-ions, cf. Fig. \ref{fig:sketch}(a). In contrast, valence electrons are less localized. In addition to the bound states (core, valence and conduction bands), if ionization processes take place, the ionization continuum (IC, cf. Fig. \ref{fig:sketch}(a)) for electronic states above the ionization edge must be taken into account. The corresponding field operators for electrons can be expanded into a complete set $\{\alpha\}$ of (i) core, (ii) valence, conduction and (iii) ionization continuum states: $\hat{\Psi}^{(\dagger)}(\mathbf{r},t)=\sum_{\alpha}\Psi^{(*)}_{\alpha}(\mathbf{r})a^{(\dagger)}_{\alpha}(t)$ with quantum number $\alpha$, single particle wave functions $\Psi_{\alpha}(\mathbf{r})$ and the fermionic annihilation (creation) operators $a^{(\dagger)}_{\alpha}$. The Hamiltonian contains the single electron energies $E_{\alpha}$ in the lattice periodic potential, the carrier-carrier interaction with Coulomb matrix element $V^{\alpha\beta}_{\alpha'\beta'}$ and the light-matter interaction $\Omega_{\alpha\alpha'}$ in second quantization:
\begin{align}
H&=\sum_{\alpha} E_{\alpha} a^{\dagger}_{\alpha}a^{\mathstrut}_{\alpha} - \hbar\sum_{\alpha,\alpha'} \Omega_{\alpha\alpha'} ~ a^{\dagger}_{\alpha}a^{\mathstrut}_{\alpha'} \nonumber \\
&+\frac{1}{2}\sum_{\alpha,\alpha',\beta,\beta'} V^{\alpha\beta}_{\alpha'\beta'} ~ a^{\dagger}_{\alpha}a^{\dagger}_{\beta}a^{\mathstrut}_{\beta'}a^{\mathstrut}_{\alpha'} \;. \label{eq:Hallg}
\end{align}
The light-matter interaction, described in length gauge discussed in the supplemental material (SM Sec. I), is given by the Rabi frequency $\Omega_{\alpha\alpha'}$, and reads $\Omega_{\alpha \alpha'}=e_0\langle \Psi_{\alpha}|\mathbf{r}\cdot\mathbf{E}(\mathbf{r},t)|\Psi_{\alpha'}\rangle/\hbar$, where $e_0$ is the elementary charge. The Coulomb matrix element reads $ V^{\alpha\beta}_{\alpha'\beta'}=\langle\Psi_{\alpha}\Psi_{\beta}|V(\mathbf{r-r'})|\Psi_{\beta'}\Psi_{\alpha'}\rangle$ with the Coulomb potential $V(\mathbf{r-r'})$ in real space discussed in the supplemental material. The single-particle eigenbasis $\Psi_{\alpha}(\mathbf{r})$ of the one-particle Hamiltonian and the corresponding eigenenergies $E_{\alpha}$ are obtained from the Schr\"odinger equation
\begin{align}
H_0\Psi_{\alpha}(\mathbf{r})=\left(-\frac{\hbar^2\nabla^2}{2m}+U(\mathbf{r})\right)\Psi_{\alpha}(\mathbf{r})=E_{\alpha}\Psi_{\alpha}(\mathbf{r}) \label{eq:SG}
\end{align}
describing electrons in a lattice periodic potential $U(\mathbf{r})$ of the ions: $U(\mathbf{r}-\mathbf{R}_{\parallel})=U(\mathbf{r})$ with a lattice vector $\mathbf{R}_{\parallel}$ in the in-plane motion of the atomically thin two-dimensional structure. The vector $\mathbf{r}$ of the electron is still three-dimensional to have access to the orbitals extending perpendicular to the plane. Eigenfunctions of the lattice periodic Hamiltonian have to fulfill the Bloch condition $\Psi(\mathbf{r+R_{\parallel}})=\exp(i\mathbf{k_{\parallel}\cdot R_{\parallel}})\Psi(\mathbf{r})$. Therefore, the eigenvectors $\Psi_{\alpha}$ and eigenenergies $E_{\alpha}$, are classified by two quantum numbers $\{\alpha\}=\{\kp,\lambda\}$, namely wave vector $\mathbf{k}_{\parallel}$ in the plane of the two-dimensional material and band index $\lambda$. The normalized eigenfunctions in the lattice periodic potential are Bloch waves \cite{Bloch1929,Wang2018,SchaferBuch}
\begin{align}
\Psi_{\lambda,\mathbf{k}_{\parallel}}(\mathbf{r})=\frac{1}{\sqrt{A l_z}}e^{i\kp\cdot\mathbf{r}_{\parallel}}u_{\lambda,\mathbf{k}_{\parallel}}(\mathbf{r}) \label{eq:Bloch}
\end{align}
with lattice periodic Bloch factor $u_{\lambda,\mathbf{k}_{\parallel}}(\mathbf{r})$, the sample area $A$ and length of the quantization volume in perpendicular direction to the material $l_z$ (sample volume $Al_z$). The lattice periodic function $u_{\lambda,\mathbf{k}_{\parallel}}(\mathbf{r})$ are orthonormalized on the in-plane unit cell: $\langle u_{\lambda,\kp}|u_{\lambda',\kp}\rangle_{UC}=V_{UC}\delta_{\lambda,\lambda'}$, with unit cell volume $V_{UC}$. Figure \ref{fig:unitCell} represents a conceptional sketch of a two-dimensional crystal built by consecutive identical unit cells. Equation \eqref{eq:Bloch} represents solutions for all states $\alpha$ -- (i) core, (ii) valence/conduction band and (iii) ionization continuum state. This accounts for the translation invariance of the crystal, a property which is also existing for vanishing coupling among the electrons in the core states of different atoms. In the following we discuss the properties of the different wave functions and the approximations made in the following separately:
\begin{figure}[t]
\begin{center}
\includegraphics[width=\linewidth]{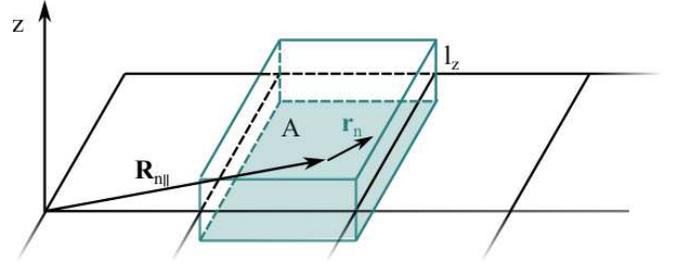}
\end{center}
\caption{Sketch of a two-dimensional crystal built by a series of unit cells, highlighted in green, with an area $A$ and thickness $l_z$. The space vector can be decomposed into a lattice vector $\mathbf{R}_{n\parallel}$ pointing towards the n-th unit cell and a unit cell local vector $\mathbf{r}_n$. The complete crystal can then be expressed by a sum over the number of repeating unit cells.}
\label{fig:unitCell}
\end{figure}

(i) As solution of the Schr\"odinger equation, Eq. \eqref{eq:Bloch} can be used to describe the core state wave functions. However, core states are strongly localized at the atomic site and it is useful to built a localized wave packet as superposition of Bloch waves of different $\mathbf{k}_{\parallel}$ to use Wannier functions \cite{Wannier1937,Marzari2012}
\begin{align}
w_{\lambda}(\mathbf{r},\mathbf{R}_{n\parallel})=\sum_{\mathbf{k}_{\parallel}}\exp(-i\mathbf{k}_{\parallel}\cdot\mathbf{R}_{n\parallel})\Psi_{\lambda,\mathbf{k}_{\parallel}}(\mathbf{r})/\sqrt{N} \label{eq:Wannier}
\end{align}
as alternative representation to Eq. \eqref{eq:Bloch} for Bloch states. The number of unit cells building the crystal is denoted by $N$. Wannier functions are essentially a real-space representation of localized orbitals and provide an extension of the concept of atomic orbitals into solids. They have assigned as quantum numbers the lattice vectors $\mathbf{R}_{n\parallel}$ of the $n$th cell, where the orbital is localized, and the band index $\lambda$. Using Eq. \eqref{eq:Bloch} an expression for the Bloch factor $u_{\lambda,\mathbf{k}_{\parallel}}(\mathbf{r})$ as a function of the Wannier orbitals $w_{\lambda}(\mathbf{r},\mathbf{R}_{n\parallel})$ reads: $u_{\lambda,\mathbf{k}_{\parallel}}(\mathbf{r})=\sqrt{V_{UC}}\sum_{\mathbf{R}_{n\parallel}}e^{-i\mathbf{k}_{\parallel}\cdot (\mathbf{r}-\mathbf{R}_{n\parallel})} w_{\lambda}(\mathbf{r},\mathbf{R}_{n\parallel})$. Since core electronic wave functions of neighbored atoms do barely overlap and electron hopping between neighboring sites is unlikely, the band dispersion is small and can be treated as a flat band. This band is built up from all the core levels of the different atoms of the crystal to account for the translation invariance. Therefore, irradiating a solid state should not be seen as light being absorbed by one individual atom with a core-hole that sits exactly at the individual atom but as a superposition of all core states represented by a collective flat band excitation.

(ii) For valence and conduction band wave functions, typically, a full band calculation needs to be performed \cite{Kohn1965,Hybertsen1984,Haastrup2018} and we will use the full Bloch equation structure Eq. \eqref{eq:Bloch} to determine the interaction matrix element $\Omega_{\alpha\alpha'}$ and $V^{\alpha\beta}_{\alpha'\beta'}$. These calculation yields also an expression for the single-particle band structure $E_{\lambda,\kp}$. At this point, the description of this work can be connected to \textit{ab initio} electronic structure theory.

(iii) The wave function of the ionization continuum can, to a good approximation, be expressed by plane waves in vacuum orthogonalized to core and band states. The orthogonalization ensures that all states, including core, valence and conduction band, and ionization continuum form a complete orthogonal basis. For the ionization continuum, where the plane wave has been orthogonalized to the Bloch wave functions we  exploite the Gram-Schmidt process\cite{Herring1940,Callaway1955,Woodruff1957}
\begin{align}
\Psi_{k_z,\kp}(\mathbf{r}) &= \frac{1}{\sqrt{V}} e^{i\mathbf{k\cdot r}} - \sum_{\lambda} \eta_{\lambda,\mathbf{k}} \Psi_{\lambda,\mathbf{k}_{\parallel}}(\mathbf{r}) \label{eq:freeWF}
\end{align}
with orthogonalization coefficient $\eta_{\lambda,\mathbf{k}} = \int_{-\infty}^{\infty} d^3r ~ \exp(i\mathbf{k\cdot r}) \Psi^{*}_{\lambda,\mathbf{k}_{\parallel}}(\mathbf{r}) /\sqrt{V}$. The ionization continuum $E_{k_z,\kp}=\hbar^2\mathbf{k}^2/2m_0+E_{Ion}$ constitute the manifold of three-dimensional parabolas with energies starting at the ionization energy $E_{Ion}$ of the material.

In principle, Eq. \eqref{eq:Bloch} (alternatively using Eq. \eqref{eq:Wannier}) and Eq. \eqref{eq:freeWF} constitute a complete set of orthonormal basis functions. In the basis of the Bloch functions Eq. \eqref{eq:Bloch}, the Hamiltonian Eq. \eqref{eq:Hallg} reads
\begin{align}
H&=\sum_{\lambda,\kp} E_{\lambda,\kp} ~ a^{\dagger}_{\lambda,\kp}a^{\mathstrut}_{\lambda,\kp} \nonumber \\
&\hspace{-3mm}+\frac{1}{2}\sum_{\substack{\lambda,\nu,\lambda',\nu' \\ \kp,\mathbf{q}_{\parallel},\kp',\mathbf{q}_{\parallel}'}} V^{\lambda\nu\nu'\lambda'}_{\kp\mathbf{q}_{\parallel}\mathbf{q}_{\parallel}'\kp'} ~ a^{\dagger}_{\lambda,\kp}a^{\dagger}_{\nu,\mathbf{q}_{\parallel}}a^{\mathstrut}_{\nu',\mathbf{q}_{\parallel}'}a^{\mathstrut}_{\lambda',\kp'} \nonumber \\
&\hspace{-3mm}- \sum_{\lambda,\lambda',\kp,\kp',\qp} \mathbf{d}^{\lambda\lambda'}_{\kp,\kp'}(\boldsymbol{Q}_{\parallel})\cdot\mathbf{E}_{-\boldsymbol{Q}_{\parallel}}(z_0,t) ~ a^{\dagger}_{\lambda,\kp}a^{\mathstrut}_{\lambda',\kp'} \;, \label{eq:Hsolid}
\end{align}
where the $\lambda$ and $\nu$-sums include all three types of states -- core, valence, conduction band and ionization continuum. Note that we use the out-of-plane wave vector component $k_z$ as quantum number for the three-dimensional states above the ionization threshold: In case that the band index $\lambda$ describes an unbound state we write $\lambda\rightarrow k_z$, which is conform with the notation used for the wave function Eq. \eqref{eq:freeWF}. 

The Coulomb interaction in Eq. \eqref{eq:Hsolid} couples two initial $(\lambda,\kp)$ and $(\nu,\mathbf{q}_{\parallel})$ with two final electron states $(\lambda',\kp')$ and $(\nu',\mathbf{q}_{\parallel}')$. The corresponding matrix element is defined by
\begin{align}
V^{\lambda\nu\nu'\lambda'}_{\kp\mathbf{q}_{\parallel}\mathbf{q}_{\parallel}'\kp'} = \langle \Psi_{\lambda,\kp}\Psi_{\nu,\mathbf{q}_{\parallel}} \mid V(\mathbf{r-r'}) \mid \Psi_{\nu',\mathbf{q}_{\parallel}'}\Psi_{\lambda',\kp'} \rangle
\end{align}
with the Coulomb potential $V(\mathbf{r-r'})$. 

In Eq. \eqref{eq:Hsolid}, for the light-matter interaction at the sample position the in-plane Fourier transform of the X-ray electric field $\mathbf{E}(\mathbf{r},t)=\sum_{\boldsymbol{Q}_{\parallel}} \mathbf{E}_{\boldsymbol{Q}_{\parallel}}(z_0,t)\exp(-i\boldsymbol{Q}_{\parallel}\cdot\mathbf{r}_{\parallel})$, the wave vector of the X-ray radiation $\boldsymbol{Q}_{\parallel}$ and $z_0$ as sheet position are introduced. Since we investigate two-dimensional materials we assume an atomically thin sheet lying at $z_0$ that the spatial variation of the electric field in $z$-direction is negligible. This approximation can be discussed in more detail: For two-dimensional materials the layer thickness lies in the range of the lattice constant. For many materials such as graphene, TMDCs or antimonene the thickness lies between \unit[0.3]{nm} and \unit[0.6]{nm}. Here, we focus on soft X-ray radiation up to an energy of \unit[500]{eV} corresponding to a wavelength of \unit[2.5]{nm} justifying the neglect of the spatial variation of the electric field perpendicular to the plane. Without the loss of generality, we set $z_0=0$ at the coordinate center. The dipole matrix element reads
\begin{align}
\mathbf{d}^{\lambda\lambda'}_{\kp,\kp'}(\boldsymbol{Q}_{\parallel})=e_0\langle \Psi_{\lambda,\kp}\mid\mathbf{r} ~ e^{i\boldsymbol{Q}_{\parallel}\cdot\mathbf{r}_{\parallel}} \mid\Psi_{\lambda',\kp'}\rangle \;, \label{eq:d}
\end{align}
which carries also the spatial in-plane component of the X-ray field. To evaluate the dynamics of X-ray electron interactions, we need to determine the interaction matrix element $\mathbf{d}^{\lambda\lambda'}_{\kp,\kp'}(\boldsymbol{Q}_{\parallel})$. While the Hamiltonian Eq. \eqref{eq:Hsolid} contains all possible electronic transitions, we discuss at this point the matrix elements for the XANES and EXAFS processes separately. We start with the XANES process in Sec. \ref{sec:xanes} and discuss the EXAFS transition in Sec \ref{sec:exafs}. The corresponding transitions are sketched in Fig. \ref{fig:sketch}(a).

\subsubsection{XANES matrix element} \label{sec:xanes}
In XANES the X-rays have sufficient energy to excite electrons from core band states to unoccupied excited states below the ionization threshold (CB-states), cf. Fig. \ref{fig:sketch}(a). This process gives rise to sudden absorption edges in the XAS spectrum. To calculate the XANES matrix element, we restrict the quantum number $\lambda$ to bands energetically below the ionization threshold and use the wave functions Eq. \eqref{eq:Bloch} in the form of the Bloch functions for the definition of the dipole matrix element $\mathbf{d}^{\lambda\lambda'}_{\kp,\kp'}(\boldsymbol{Q}_{\parallel})$, Eq. \eqref{eq:d}. This implies a momentum selection rule of $\mathbf{k}'_{\parallel}=\mathbf{k}_{\parallel}-\boldsymbol{Q}_{\parallel}+\mathbf{G}_{\parallel}$ for the optical transition, where the momentum is conserved up to a reciprocal lattice vector $\mathbf{G}_{\parallel}$ and detailed in the SM Sec. II. The wave vector $\kp$ is a reduced wave vector lying in the first Brillouin zone. Already soft X-rays have a wave number of tens of percent of the Brillouin zone. Therefore, when adding $\kp$ and $\boldsymbol{Q}_{\parallel}$ the resultant vector can lie outside of the first Brillouin zone. The reciprocal lattice vector $\gp\neq 0$ accounts for Umklapp processes, refolding the resultant vector to its equivalent wave vector in the first Brillouin zone \cite{MadelungBuch}. A detailed derivation of the XANES dipole matrix operator acting partly as a derivative on $a^{\dagger}_{\lambda,\kp}a^{\mathstrut}_{\lambda',\kp'}$ is provided in the SM. Up to this point, coming from Eq. \eqref{eq:Hsolid}, the X-ray wave vector is defined by a general Fourier transform over the complete space. Therefore, to be consistent with the definition of the electronic wave vector $\kp$ we replace $\boldsymbol{Q}_{\parallel}\rightarrow\mathbf{Q}_{\parallel}+\gp$, where $\mathbf{Q}_{\parallel}$ is now defined within the first Brillouin zone. We obtain 
\begin{widetext}
\begin{align} 
\mathbf{X}^{\lambda\lambda'}_{\mathbf{k}_{\parallel}+\mathbf{Q}_{\parallel},\mathbf{k}_{\parallel}}(\mathbf{G}_{\parallel})&=-\frac{ie_0}{V_{UC}}  \langle
\begin{pmatrix}
\nabla_{\kp+\mathbf{Q_{\parallel}}} \\ iz
\end{pmatrix}
u_{\lambda,\kp+\mathbf{Q}_{\parallel}} \mid e^{i\mathbf{G}_{\parallel}\cdot\mathbf{r}_{\parallel}} \mid u_{\lambda',\kp} \rangle -  ie_0\delta_{\lambda,\lambda'} \nabla_{\kp+\mathbf{Q_{\parallel}}}  \;. \label{eq:XANES}
\end{align}
\end{widetext}
To clarify that the matrix operator Eq. \eqref{eq:XANES} describes only X-ray induced transitions with initial and final electronic state below the ionization threshold (\textbf{X}ANES) from now on the labeling $\mathbf{d}^{\lambda\lambda'}_{\mathbf{k}_{\parallel}+\mathbf{Q}_{\parallel},\mathbf{k}_{\parallel}}(\mathbf{G}_{\parallel})\rightarrow\mathbf{X}^{\lambda\lambda'}_{\mathbf{k}_{\parallel}+\mathbf{Q}_{\parallel},\mathbf{k}_{\parallel}}(\mathbf{G}_{\parallel})$ is chosen. In Eq. \eqref{eq:XANES}, the first term describes interband transitions between core and unoccupied conduction band states. XANES transitions are labelled K, L, M, etc. depending on the principle quantum number of the initial core band $\lambda$, based on the Siegbahn notation \cite{Siegbahn1916,Jenkins1991}. For example, K transitions involve 1$s$ electrons, while excitation of 2$s$ and 2$p$ electrons occurs at L edges. The latter, as energetically following edges, are divided into L$_1$ for 2$s$ electrons and L$_2$, L$_3$ for 2$p$ electrons (total angular momentum $J=1/2$ and $J=3/2$). The  momentum conservation accounts for a transfer of the in-plane field momentum to the optically excited electron. The second term with the momentum gradient acting on the creation operator in Eq. \eqref{eq:Hsolid}, leads to a wave vector gradient, which changes according to the acceleration theorem with a rate proportional to the X-ray field \cite{Bloch1929} and describes the coupling strength of X-ray excitation to intraband transitions \cite{Golde2008,Song2020}. Since X-ray radiation includes excitation energies from a hundred of eV up to tens of keV, the intraband dynamics do not significantly change the optical response in a rotating wave approximation. 

Note, that the Bloch factors of the core states in Eq. \eqref{eq:XANES} can also be expressed as functions of the Wannier orbitals Eq. \eqref{eq:Wannier} if this representation is desired.

\subsubsection{EXAFS matrix element} \label{sec:exafs}
After having discussed the XANES dipole matrix element we turn our attention to the EXAFS matrix element between a Bloch-like core state $\{\kp,\lambda\}$, Eq. \eqref{eq:Bloch}, and ionization continuum (IC) state $\{\kp,\lambda'\equiv k_z\}$, cf. Fig. \ref{fig:sketch}. Inserting the orthogonalized plane waves Eq. \eqref{eq:freeWF} and the Bloch band states Eq. \eqref{eq:Bloch} into the definition of the dipole matrix element Eq. \eqref{eq:d} yields formally
\begin{align}
\mathbf{Y}^{\lambda k_z}_{\mathbf{k}_{\parallel},\kp'}(\boldsymbol{Q}_{\parallel})&=e_0 \langle\Psi_{\lambda,\mathbf{k}_{\parallel}} \mid \mathbf{r} ~ e^{i\boldsymbol{Q_{\parallel}\cdot r_{\parallel}}} \mid \mathbf{k}'\rangle \nonumber \\
&+ e_0\sum_{\lambda'} \eta_{\lambda',\mathbf{k}'} \langle\Psi_{\lambda,\mathbf{k}_{\parallel}} \mid \mathbf{r} ~ e^{i\boldsymbol{Q_{\parallel}\cdot r_{\parallel}}} \mid \Psi_{\lambda',\mathbf{k'}_{\parallel}} \rangle \label{eq:exafsallg}
\end{align}
with the notation $\langle\mathbf{r}|\mathbf{k}\rangle=\exp(i\mathbf{k\cdot r})/\sqrt{V}$ resulting from the unorthogonalized plane wave character of the final state. The second contribution originates from the orthogonalization coefficients defined below Eq. \eqref{eq:freeWF}. To clarify that the dipole matrix element now describes transitions into the ionization continuum it is denoted by $\mathbf{d}^{\lambda k_z}_{\mathbf{k}_{\parallel},\kp'}(\boldsymbol{Q}_{\parallel})\rightarrow\mathbf{Y}^{\lambda k_z}_{\mathbf{k}_{\parallel},\kp'}(\boldsymbol{Q}_{\parallel})$. The EXAFS matrix element Eq. \eqref{eq:exafsallg} can be calculated in similar way as the XANES matrix element, however the final state wave vector is, in contrast to the initial state, three-dimensional. After some manipulations, detailed in the SM Sec. III, and restricting $\boldsymbol{Q}_{\parallel}$ to the first Brillouin zone the EXAFS matrix element is obtained as
\begin{align}
\mathbf{Y}^{\lambda k_z}_{\kp+\qp,\kp}(\gp)&= \sum_{\lambda'}\eta_{\lambda',\mathbf{k}}\mathbf{X}^{\lambda\lambda'(\text{inter})}_{\kp+\qp,\kp} \nonumber \\
&\hspace{-20mm}-\frac{ie_0}{\sqrt{V_{UC}}} \langle 
\begin{pmatrix}
\nabla_{\kp+\mathbf{Q}_{\parallel}} \\
iz
\end{pmatrix}
u_{\lambda,\kp+\qp} \mid e^{i\mathbf{G}_{\parallel}\cdot\mathbf{r}_{\parallel}} e^{ik_z z} \rangle \; . \label{eq:Hexafs}
\end{align}
Similar to Eq. \eqref{eq:XANES} (XANES), we find an interband term describing the optical transition between a Bloch and a plane wave state (second line). Note that we consider for EXAFS only transitions from the material into the ionization continuum. Therefore the matrix element Eq. \eqref{eq:Hexafs} contains no intraband interaction because of the forced orthogonality of the ionization continuum to all band states. Consequently, the appearing XANES matrix element in Eq. \eqref{eq:Hexafs}, stemming from the orthogonalization, is restricted to its interband part. In principle, also an acceleration of the free electrons is possible and a third contribution to the matrix element $\mathbf{Y}^{k_z k_z}_{\kp+\qp,\kp}(\gp)$ is present, which is proportional to the X-ray field times the momentum gradient of the electronic creation operator \cite{Golde2008}. However, since this is no EXAFS transition, goes beyond our description, and is not visible in the absorption spectrum this contribution is neglected.

\subsubsection{Hamiltonian, discriminating XANES and EXAFS processes}

Using the calculated matrix elements for the XANES and EXAFS processes, the Hamiltonian containing the electron dispersion and the interaction of X-ray electric field with the crystal electrons reads
\begin{align}
H&=\sum_{\lambda,\mathbf{k}_{\parallel}} E_{\lambda,\mathbf{k}_{\parallel}} ~ a^{\dagger}_{\lambda,\mathbf{k}_{\parallel}}a^{\mathstrut}_{\lambda,\mathbf{k}_{\parallel}} + \sum_{\kp,k_z} E_{k_z,\mathbf{k}_{\parallel}} ~ a^{\dagger}_{k_z,\kp}a^{\mathstrut}_{k_z,\kp} \nonumber \\
&\hspace{-5mm}- \sum_{\substack{\lambda,\lambda' \\ \mathbf{k}_{\parallel},\mathbf{Q}_{\parallel},\gp}} \mathbf{X}^{\lambda\lambda'}_{\mathbf{k}_{\parallel}+\mathbf{Q}_{\parallel},\mathbf{k}_{\parallel}}(\gp)\cdot\mathbf{E}_{-\mathbf{Q}_{\parallel}+\gp}(t) ~ a^{\dagger}_{\lambda,\mathbf{k}_{\parallel}+\mathbf{Q}_{\parallel}}a^{\mathstrut}_{\lambda',\mathbf{k}_{\parallel}} \nonumber \\
&\hspace{-5mm}- \sum_{\substack{\lambda,k_z \\ \kp,\mathbf{Q}_{\parallel},\gp}} \left( \mathbf{Y}^{\lambda k_z}_{\mathbf{k}_{\parallel}+\mathbf{Q}_{\parallel},\kp}(\gp)\cdot\mathbf{E}_{-\mathbf{Q}_{\parallel}+\gp}(t) ~ a^{\dagger}_{\lambda,\mathbf{k}_{\parallel}+\mathbf{Q}_{\parallel}}a^{\mathstrut}_{k_z,\kp} \right. \nonumber \\
&\left.+ \text{H.c.} \right) \nonumber \\
&\hspace{-5mm}+\frac{1}{2}\sum_{\substack{\lambda,\nu,\lambda',\nu' \\ \kp,\mathbf{q}_{\parallel},\kp',\mathbf{q}_{\parallel}'}} V^{\lambda\nu\nu'\lambda'}_{\kp\mathbf{q}_{\parallel}\mathbf{q}_{\parallel}'\kp'} ~ a^{\dagger}_{\lambda,\kp}a^{\dagger}_{\nu,\mathbf{q}_{\parallel}}a^{\mathstrut}_{\nu',\mathbf{q}_{\parallel}'}a^{\mathstrut}_{\lambda',\kp'} \;. \label{eq:Hint}
\end{align}
The unpertubed single particle energies in a lattice periodic atomic potential are described by the first line of Eq. \eqref{eq:Hint}. The crystal band index $\lambda$ in the first term incorporates all bound bands, starting at the core level 1$s$ state up to the comparable delocalized valence and conduction band states. The second line describes XANES transitions between two-dimensional core and conduction band states. The third line describes EXAFS transitions between two-dimensional initial and three-dimensional final states lying above the ionization threshold of the crystal. It should be remembered, that $k_z$ describes the out-of-plane wave vector and acts as quantum number for the unbound states in a three-dimensional continuum. The last line describes Coulomb interaction between the carriers. It couples two initial states $(\lambda,\kp)$, $(\nu,\mathbf{q}_{\parallel})$ and two final states $(\lambda',\kp')$, $(\nu',\mathbf{q}_{\parallel}')$. Computing the Coulomb matrix element as is shown in the SM we can also derive momentum selection rules holding during Coulomb scattering. In detail, we can use $\sum_{\gp}\delta_{\kp,\kp'+\mathbf{p}_{\parallel}-\gp}$ and $\sum_{\gp'}\delta_{\mathbf{q}_{\parallel},\mathbf{q}_{\parallel}'-\mathbf{p}_{\parallel}-\gp'}$, where $\mathbf{p}_{\parallel}$ describes the momentum transfer during the carrier-carrier interaction and the reciprocal lattice vectors $\gp$ and $\gp'$ takes into account for possible Umklapp processes. The Coulomb selection rules can be inserted in the Hamiltonian Eq. \eqref{eq:Hint} if Coulomb scattering wants to be evaluated explicitly.

\subsection{Dipole matrix elements in tight binding approximation}
The character of the dipole transitions, Eq. \eqref{eq:XANES} and Eq. \eqref{eq:Hexafs}, including the lattice geometry, which is of essential importance for X-ray experiments in the solid state, as well as the orbital band composition are encoded in the lattice periodic function $u_{\lambda,\kp}(\mathbf{r})$. Access to the Bloch functions (or Wannier functions) typically requires computationally expensive methods. To evaluate explicitly the matrix element and to obtain more analytical insights into X-ray induced electronic transitions we use the tight binding method \cite{Jones1934,Slater1954}, in chemistry usually referred to as H\"uckel theory \cite{Huckel1931}, for core and conduction band wave function. The method consists of the assumption to approximate the Wannier functions with atomic orbitals. The Bloch states $\Psi_{\lambda,\mathbf{k}_{\parallel}}(\mathbf{r})$ are then expanded in terms of a linear combination of atomic orbitals $\langle\mathbf{r}|\lambda,\beta,j,0\rangle=\phi^{\lambda}_{\beta,j}(\mathbf{r})$ (orbital of type $j$ on atom $\beta$) of the composing atoms, which are eigenstates of the single atom Hamiltonian:
\begin{align}
\Psi_{\lambda,\mathbf{k}_{\parallel}}(\mathbf{r}) = \frac{1}{\sqrt{N}} \sum_{\beta,j,\mathbf{R}_{\beta\parallel}} C_{\beta j,\mathbf{k}_{\parallel}}^{\lambda}  e^{i\mathbf{k}_{\parallel}\cdot\mathbf{R}_{\beta\parallel}} \phi^{\lambda}_{\beta,j}(\mathbf{r}-\mathbf{R}_{\beta\parallel}). \label{eq:TightBind} 
\end{align}
The atomic orbitals spatially decay with a typical constant of $\zeta=Z/a_B$, where $a_B$ denotes the hydrogen Bohr radius and $Z$ is the effective nuclear charge, incorporating electron screening effects. Values for an effective nuclear charge $Z$ in atoms or ions is provided by the Slater rule \cite{Slater1930}, which can be used as a first estimate also in crystals. For more accurate values first-principle calculations are necessary. In Eq. \eqref{eq:TightBind}, the coordinates of the atoms in the crystal lattice are denoted by $\mathbf{R}_{\beta\parallel}$ and the tight binding coefficients $C^{\lambda}_{\beta j,\mathbf{k}_{\parallel}}$ determine the weight of the different orbitals. For core states the atomic orbitals decay rapidly from the atomic position that their overlap is negligible small. In the following we revisit the XANES and EXAFS matrix elements exploiting the wave function Eq. \eqref{eq:TightBind} for core and conduction band wave function and derive the solid state electron-X-ray Hamiltonian in tight binding approximation.

\subsubsection{XANES}
The XANES matrix element is evaluated by inserting the tight binding wave function Eq. \eqref{eq:TightBind} for initial and final state into the formal definition Eq. \eqref{eq:d}. Expanding the integral into a sum over unit cells at the lattice vector $\mathbf{R}_{\beta\parallel}$ leads to the matrix element
\begin{widetext}
\begin{align}
\mathbf{X}^{\lambda\lambda'}_{\mathbf{k}_{\parallel}+\mathbf{Q}_{\parallel},\mathbf{k}_{\parallel}}(\gp)&=-e_0\sum_{\alpha,\beta,i,j}\sum_{\boldsymbol{\delta}_{\beta\alpha}} C^{*\lambda}_{\beta j,\mathbf{k}_{\parallel}+\mathbf{Q}_{\parallel}}C^{\lambda'}_{\alpha i,\mathbf{k}_{\parallel}} e^{i\mathbf{k_{\parallel}}\cdot\boldsymbol{\delta}_{\beta\alpha}} \langle \lambda,\beta,j,0 \mid \mathbf{r} ~ e^{i(\mathbf{Q}_{\parallel}+\gp)\cdot \mathbf{r}_{\parallel}} \mid \lambda',\alpha,i,\boldsymbol{\delta}_{\beta\alpha}\rangle \nonumber \\
&-i e_0\sum_{\alpha,\beta,i,j} \sum_{\boldsymbol{\delta}_{\beta\alpha}} C^{*\lambda}_{\beta j,\mathbf{k}_{\parallel}+\qp}C^{\lambda'}_{\alpha i,\mathbf{k}_{\parallel}} e^{i\mathbf{k}_{\parallel}\cdot\boldsymbol{\delta}_{\beta\alpha}} \langle \lambda,\beta,j,0 \mid  e^{i(\mathbf{Q}_{\parallel}+\gp)\cdot \mathbf{r}_{\parallel}} \mid \lambda',\alpha,i,\boldsymbol{\delta}_{\beta\alpha}\rangle \nabla_{\mathbf{k}_{\parallel}+\mathbf{Q}_{\parallel}}  \label{eq:XANEStb}
\end{align}
\end{widetext}
including the same momentum selection rule as previously, Eq. \eqref{eq:XANES} and using that $\qp$ is restricted to the first Brillouin zone. The projection onto atomic orbital basis yields $\langle\mathbf{r} | \lambda,\beta,j,\boldsymbol{\delta}_{\beta\alpha}\rangle=\phi_{\beta j}^{\lambda}(\mathbf{r}-\boldsymbol{\delta}_{\beta\alpha})$. The electron momentum-dependence is carried out by the tight binding coefficients and the phase factor $\exp(i\mathbf{k}_{\parallel}\cdot\boldsymbol{\delta}_{\beta\alpha})$, where $\boldsymbol{\delta}_{\beta\alpha}=\mathbf{R}_{\beta\parallel}-\mathbf{R}_{\alpha\parallel}$ stands for the next-neighbor vectors connecting the atoms. The sum over the sublattices $\alpha$, $\beta$ includes neighboring atoms of arbitrary order. Since we investigate the interaction of strongly localized core electrons, their spatial extension decays rapidly even compared to the wavelength of soft and medium X-rays up to an energy of \unit[$\sim 4$-5]{keV}. Therefore, it is reasonable to treat the transition integral in dipole approximation, i.e. perform an expansion of the radiation field in zeroth order $\exp(i(\qp+\gp)\cdot\mathbf{r}_{\parallel})\approx 1$. We have carefully checked, see SM Sec. IV, that the inclusion of the $\qp$-dependence to the transition integral does not change the result at this point. Since the spatial localization of the core electrons increases with the atomic weight, the dipole approximation becomes better the heavier the constituting atoms are. Treating the transition integral within the dipole approximation, the optical selection rules known from atomic spectroscopy are recovered. However, thanks to the tight binding wave function of the crystalline solid and the plane wave decomposition in Eq. \eqref{eq:XANEStb} we include solid state properties, in particular the lattice periodicity. Because the core orbitals are generally more localized compared to other states the optical transitions involving core states are generally weaker than transitions between valence and conduction band adressed in optical experiments with visible light. Applying the dipole approximation to the second line of  Eq. \eqref{eq:XANEStb} the integral turns into an overlap integral $\langle \lambda,\beta,j,0 \mid \lambda',\alpha,i,\boldsymbol{\delta}_{\beta\alpha}\rangle$. The overlap is generally small and we may assume that the chosen orbitals are orthogonal to each other showing that the second line in Eq. \eqref{eq:XANEStb} describes intraband transitions similar to Eq. \eqref{eq:XANES}.

\subsubsection{EXAFS}
As for the XANES transitions, we evaluate the EXAFS matrix element for tight binding wave functions. The starting point is the EXAFS dipole matrix element Eq. \eqref{eq:exafsallg}, now with the initial state electronic wave function $\Psi_{\lambda,\kp}(\mathbf{r})$ in the tight binding approach. First we investigate the orthogonalization coefficients defined below Eq. \eqref{eq:freeWF}, which appears as a sum involving all bound bands. With the main $n$, angular $l$ and magnetic $m$ quantum numbers, which determine the atomic orbitals, the coefficients can be calculated to be $\eta_{\lambda,\mathbf{k}}=i^l\sqrt{(2\zeta)^{2n+1}/(2n)!}J_{n,l}(k)Y_{lm}(\vartheta_k,\varphi_k)/2\pi^2$ with the spherical harmonics $Y_{lm}$ and the radial function $J_{n,l}(k)=\sqrt{\pi/2k}(\zeta^2+k^2)^{-(n+1)/2}\Gamma(n-1+l)P^{-l}_n[\zeta(\zeta^2+l^2)^{-1/2}]$ with the associated Legendre polynomials $P_n^l(x)$. The orthogonalization coefficient to the 1$s$ state is proportional to $\eta_{1s,\mathbf{k}}\propto (\zeta^2+k^2)^{-2}$. Here, $\eta_{1s,\mathbf{k}}$ decreases to a value of \unit[10]{\%} at approximately $k=\sqrt{2}\zeta$. Generally, because the 1$s$-orbital has the smallest extent in real space, its orthogonalization contribution determines the strength of the orthogonalization contribution to the EXAFS dipole matrix element. Further, we see that the orthogonalization coefficient depends on the atomic number and gains importance with increasing atomic weight. With an exemplary effective atomic number of $Z=5.7$, stemming from the Slater rule for carbon \cite{Slater1930}, we obtain a wave number of \unit[140]{nm$^{-1}$} corresponding to an energy of \unit[5]{eV}. We see that the coefficients decrease rapidly with increasing energy (starting at the ionization threshold) since the orthogonalization contributes only close to the surface. Since the coefficients $\eta_{\lambda,\mathbf{k}}$ decrease rapidly to zero with increasing energy, we use in the following free electronic continuum states for simplicity and obtain for the first term of Eq. \eqref{eq:exafsallg}
\begin{align}
\mathbf{Y}^{\lambda k_z}_{\kp+\qp,\kp}(\gp)&=\frac{-e_0}{\sqrt{V_{UC}}} \sum_{\beta,j}C^{*\lambda}_{\beta j,\mathbf{k}_{\parallel}+\qp} \nonumber \\
&\times\langle \lambda,\beta,j,0 \mid \mathbf{r} \mid \mathbf{k}+\qp-\gp \rangle \;. \label{eq:PWapprox}
\end{align}
However, the plane wave approximation \eqref{eq:PWapprox} has to be investigated carefully for each material independently.

To illustrate the approach, we focus on K-shell transitions, namely $\lambda=1s$ for the initial state. The calculation of the transition integral from a 1$s$ core electron to the plane wave state is included in the SM. We obtain
\begin{align}
\mathbf{Y}^{1sk_z}_{\kp+\qp,\kp}(\gp)&=\frac{-e_0}{\sqrt{V_{UC}}} \sum_{\beta,n} C^{*1s}_{\beta 1s,\mathbf{k}_{\parallel}+\qp} \nonumber \\
&\times\frac{ 32\sqrt{\pi} i\zeta^{5/2} (\mathbf{k}+\qp-\gp)_n}{(\zeta^2+(\mathbf{k}+\qp-\gp)^2)^3} \hat{\mathbf{e}}_n  \label{eq:1sexafs}
\end{align}
with the unit vector $\hat{\mathbf{e}}_n$ in cartesian coordinates. Equation \eqref{eq:1sexafs} consists of the tight-binding coefficients $C^{*1s}_{\beta 1s,\mathbf{k}_{\parallel}+\qp}$, which carry the lattice periodicity, and the transition integral (second line), which carries the unit cell information. The magnitude and width of this form factor peaks are determined by the spatial electron distribution. Therefore, the form factor is weakened by the effective inverse Bohr radius $\zeta$, cf. Eq. \eqref{eq:1sexafs} denominator. Further, we see that the optical transition into the ionization continuum is unpolarized in contrast to transitions within the material. Therefore, the optical transition of internal states into the vacuum are independent of the incident angle of the light.
\begin{figure}[t]
\begin{center}
\includegraphics[width=\linewidth]{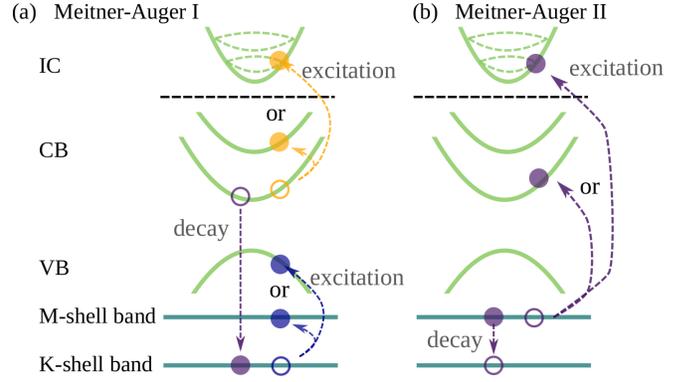}
\end{center}
\caption{Sketch of possible Meitner-Auger transitions included in the Hamiltonian. (a) Relaxation of a conduction band electron and core-hole accompanied by an excitation of a core/valence band electron (blue arrows depending on Pauli blocking) or by an excitation of conduction band electrons (yellow arrows) previously injected by the X-ray excitation. (b) Recombination of a core-hole with a core electron accompanied by an interband excitation of a core/valence electron.}
\label{fig:recomb}
\end{figure}

It is interesting to compare X-ray absorption to the complementary technique X-ray diffraction used to measure the structure of crystals. The difference between EXAFS and X-ray diffraction lies in the choice of the observable: While for the EXAFS a transmission or reflection is measured under the same angle as the X-ray incident angle, for X-ray diffraction also the signal under a different angle than the incident is measured. In particular, we have checked that the conventional X-ray diffraction of the ground state is included in our description for $\omega\rightarrow\infty$. The diffraction is determined by the Fourier transform of the electron distribution and often called form factor. As consequence the X-ray propagation involves a reduced sensitivity to light atoms with stronger localized and only weakly screened inner shell electrons\cite{Wang2020}. Moving to electron diffraction the form factor can be calculated from the X-ray form factor by the Mott-Bethe equation \cite{Mott1929,Bethe1930}, which takes additionally to the elastic scattering at the electron clouds also nucleus scattering into account. Then also crystals of lighter element can be resolved and measured.

\section{X-ray Bloch equations} \label{sec:Bloch}
The observable describing the X-ray response of the material is the detected X-ray field (Sec. \ref{sec:Absorption}), which can be calculated from the incident field interfering with the excited dipole density in the sample in reflection or transmission geometry. To derive the dipole density $\mathbf{P}(\mathbf{r},t)$ as a function of the electric field $\mathbf{E}(\mathbf{r},t)$, we start from the light-matter interaction Hamiltonian (cf. SM). From  Eq. \eqref{eq:Hsolid} we can identify the two-dimensional macroscopic polarization density
\begin{align}
\mathbf{P}^{2D}_{\mathbf{Q}_{\parallel}+\gp}(t)&= \frac{1}{A} \sum_{\lambda_1\neq\lambda_2,\kp} \mathbf{d}^{\lambda_1\lambda_2}_{\mathbf{k}_{\parallel}+\mathbf{Q}_{\parallel},\kp} p^{\lambda_1\lambda_2}_{\mathbf{k}_{\parallel}+\mathbf{Q}_{\parallel},\kp}(t) \:, \label{eq:obs}
\end{align}
which determines the X-ray response (Sec. \ref{sec:Absorption}) in wave number space. Here, we explicitly exclude the intraband contribution as discussed in Sec. \ref{sec:Hamilton1}. Please note that $\qp$ is defined within the first Brillouin zone. For the X-ray dynamics, the relevant quantities occurring in Eq. \eqref{eq:obs} are the transition $p^{\lambda_1\lambda_2}_{\mathbf{k}_{\parallel}+\mathbf{Q}_{\parallel},\kp}(t)=\langle a^{\dagger}_{\lambda_1,\mathbf{k}_{\parallel}+\mathbf{Q}_{\parallel}}a^{\mathstrut}_{\lambda_2,\kp}\rangle (t)$ defined as expectation value for a single electronic transition between the states $|\lambda_1,\kp+\mathbf{Q}_{\parallel}\rangle$ and $|\lambda_2,\kp\rangle$, which can be excited if the corresponding dipole matrix element $\mathbf{d}^{\lambda_1\lambda_2}_{\mathbf{k}_{\parallel}+\mathbf{Q}_{\parallel},\kp}$ does not vanish. Due to the spatial resolution of the X-ray light and a possibly non-orthogonal incidence, the non-diagonal character of the transitions and occupations in momentum space ($\qp\neq 0$) have to be included. The electronic transitions can be derived from the Heisenberg equation of motion using the Hamiltonian Eq. \eqref{eq:Hint}. The corresponding equation of motion for the microscopic transition reads
\begin{widetext}
\begin{align}
i\hbar\frac{d}{dt} p^{\lambda_1\lambda_2}_{\kp+\qp,\kp}&= \left(E^{\lambda_2}_{\kp}-E^{\lambda_1}_{\kp+\qp} \right)p^{\lambda_1\lambda_2}_{\kp+\qp,\kp} \nonumber \\
&-  \sum_{\lambda,\qp',\gp}\mathbf{E}_{-\qp'+\gp}(t)\cdot \left( \mathbf{d}^{\lambda_2\lambda}_{\kp,\kp-\qp'}(\gp) \sigma^{\lambda_1\lambda}_{\kp+\qp,\kp-\qp'} - \mathbf{d}^{\lambda\lambda_1}_{\kp+\qp+\qp',\kp+\qp}(\gp) \sigma^{\lambda\lambda_2}_{\kp+\qp+\qp',\kp} \right) \nonumber \\
&+\sum_{\substack{\mu,\lambda,\nu \\ \kp',\mathbf{q}_{\parallel},\mathbf{p}_{\parallel}}}  \left( V^{\lambda_2\mu\lambda\nu}_{\kp,\kp',\mathbf{q}_{\parallel},\mathbf{p}_{\parallel}} - V^{\lambda_2\mu\nu\lambda}_{\kp,\kp',\mathbf{p}_{\parallel},\mathbf{q}_{\parallel}}  \right) \sigma^{\mu\nu}_{\kp',\mathbf{p}_{\parallel}}\sigma^{\lambda_1\lambda}_{\kp+\qp,\mathbf{q}_{\parallel}} \nonumber \\
&-\sum_{\substack{\mu,\lambda,\nu \\ \kp',\mathbf{q}_{\parallel},\mathbf{p}_{\parallel} }}  \left( V^{\mu\lambda\nu\lambda_1}_{\kp',\mathbf{q}_{\parallel},\mathbf{p}_{\parallel},\kp+\qp} - V^{\lambda\mu\nu\lambda_1}_{\mathbf{q}_{\parallel},\kp',\mathbf{p}_{\parallel},\kp+\qp}  \right) \sigma^{\lambda\nu}_{\mathbf{q}_{\parallel},\mathbf{p}_{\parallel}}\sigma^{\mu\lambda_2}_{\kp,\kp} \nonumber \\
&+ i\hbar\partial_t p^{\lambda_1\lambda_2}_{\kp+\qp,\kp} \big|_{\text{coll}} \;. \label{eq:BlochGl}
\end{align}
\end{widetext}
Equation \eqref{eq:BlochGl} describes the dynamics of an X-ray induced electronic transitions $p^{\lambda_1\lambda_2}_{\kp+\qp,\kp}$ within a two-dimensional material defining the response in Eq. \eqref{eq:obs}. The first term incorporates the free oscillation of the transition with the single-particle energies of initial and final state. The second line of Eq. \eqref{eq:BlochGl} describes the coupling to the X-ray field. For now, $\sigma^{\lambda\lambda'}_{\mathbf{k}_1,\mathbf{k}_2}=\langle a^{\dagger}_{\lambda,\mathbf{k}_1}a^{\mathstrut}_{\lambda',\mathbf{k}_2} \rangle$ describes a general density matrix element. In case that $\lambda\neq\lambda'$ this expectation value describes a non-local ($\mathbf{k}_1\neq\mathbf{k}_2$) interband transition $p^{\lambda\lambda'}_{\mathbf{k}_1,\mathbf{k}_2}$. In contrast, for $\lambda=\lambda'$ the quantity stands for a non-local electron occupation $f^{\lambda}_{\mathbf{k}_1,\mathbf{k}_2}$. The two different wave number indices indicate the existence of spatial correlations in a spatially inhomogeneous system. Therefore, Eq. \eqref{eq:BlochGl} allows for the description of translational non-invariant, spatially localized, X-ray excitation of the crystal. The last three lines describe the Coulomb contribution. The many-particle interaction leads to a coupling to the dynamics of higher-order expectation values, known as hierarchy problem, which is treated by exploiting the cluster expansion \cite{Axt1994,Lindberg1994,Fricke1996}. The action of the Coulomb interaction is divided in Hartree-Fock (third and fourth line) and collision contributions $p^{\lambda_1\lambda_2}_{\kp+\qp,\kp} \big|_{\text{coll}}$ including many-body interaction and scattering-induced dephasing beyond the Hartree-Fock level \cite{Kochbuch}. Here, we see that our approach includes naturally many-body interaction in a self-consistent way: Depending on the band index combination of the Coulomb matrix element, the third and fourth line implicitly include different kinds of Coulomb interaction mechanisms. For example: i) when all band indices are equal we find a band renormalization. ii) In case of two $\lambda_1$ and two $\lambda_2$ indices we can have electron-core-hole interaction giving rise to core-excitons. iii) With three occupied and one unoccupied band indices (or vise versa) the third and fourth line describe Meitner-Auger-type interaction. Meitner-Auger transitions are recombinations of electron and core-hole accompanied by an energetic elevation of a second electron. Here, depending on the excitation conditions and band structure, one has to carefully investigate, which terms contribute to the dynamics of the X-ray induced transition. In principle, different kinds of Meitner-Auger transitions are included in the Hamiltonian. Figure \ref{fig:recomb} shows exemplary de-excitations of the core-hole. Via Coulomb coupling this interband transitions can excite either transitions within the core and valence bands (blue arrows) or excite transitions within the conduction bands (yellow arrows), cf. Fig. \ref{fig:recomb}(a). In the following we term transitions within the core/valence bands or conduction/vacuum bands as intersubband transitions to distinguish from interband transitions, where an electron changes between occupied core/valence and unoccupied conduction/vacuum bands. Since the X-ray irradiation can also excite electron occupations, discussed at the end of the section (cf. Eq. \eqref{eq:Dichte}), a core-hole recombination can also excite intersubband transitions in the conduction band as sketched in Fig. \ref{fig:recomb}(a) via Coulomb coupling. However, the core-hole can also recombine with a core-electron from another shell and excite a transition of a core-hole to the conduction bands or even out from the sample, cf. Fig. \ref{fig:recomb}(b). By inverting the arrows the complex conjugated processes are sketched, which are also included in the Hamiltonian and are called impact ionization.

For spectrally sufficient sharp X-ray pulses exciting only the transition between the bands $\lambda_1$ and $\lambda_2$ we can obtain more analytical insights from Eq. \eqref{eq:BlochGl}:
\begin{widetext}
\begin{align}
i\hbar\frac{d}{dt} p^{\lambda_1\lambda_2}_{\kp+\qp,\kp}&=\left(E^{\lambda_2}_{\kp}-E^{\lambda_1}_{\kp+\qp} \right)p^{\lambda_1\lambda_2}_{\kp+\qp,\kp} \nonumber \\
&- \sum_{\qp',\gp} \mathbf{E}_{-\mathbf{Q}'_{\parallel}+\gp}(t)\cdot \left( \mathbf{d}^{\lambda_2\lambda_1}_{\kp,\kp-\qp'}(\gp) f^{\lambda_1}_{\kp+\qp,\kp-\qp'} - \mathbf{d}^{\lambda_2\lambda_1}_{\kp+\qp+\qp',\kp+\qp}(\gp) f^{\lambda_2}_{\kp+\qp+\qp',\kp} \right)  \nonumber \\
&+ \sum_{\kp',\mathbf{q}_{\parallel},\mathbf{p}_{\parallel}} \left(V^{\lambda_1\lambda_1\lambda_1\lambda_1}_{\kp',\mathbf{q}_{\parallel},\mathbf{p}_{\parallel},\kp+\qp} f^{\lambda_1}_{\kp',\mathbf{p}_{\parallel}}p^{\lambda_1\lambda_2}_{\mathbf{q}_{\parallel},\kp} - V^{\lambda_2\lambda_2\lambda_2\lambda_2}_{\kp,\kp',\mathbf{q}_{\parallel},\mathbf{p}_{\parallel}} f^{\lambda_2}_{\kp',\mathbf{p}_{\parallel}}p^{\lambda_1\lambda_2}_{\kp+\qp,\mathbf{q}_{\parallel}} \right) \nonumber \\
&-\sum_{\kp',\mathbf{q}_{\parallel},\mathbf{p}_{\parallel}} \left( V^{\lambda_2\lambda_1\lambda_1\lambda_2}_{\kp,\kp',\mathbf{q}_{\parallel},\mathbf{p}_{\parallel}} f^{\lambda_1}_{\kp+\qp,\mathbf{q}_{\parallel}}p^{\lambda_1\lambda_2}_{\kp',\mathbf{p}_{\parallel}} - V^{\lambda_1\lambda_2\lambda_2\lambda_1}_{\kp',\mathbf{q}_{\parallel},\mathbf{p}_{\parallel},\kp+\qp} f^{\lambda_2}_{\mathbf{q}_{\parallel},\kp}p^{\lambda_1\lambda_2}_{\kp',\mathbf{p}_{\parallel}} \right) \nonumber \\
&+\sum_{\mu,\lambda,\kp',\mathbf{q}_{\parallel},\mathbf{p}_{\parallel}}\left[  V^{\lambda_2\mu\lambda\lambda_1}_{\mathbf{q}_{\parallel},\kp',\mathbf{p}_{\parallel},\kp+\qp} p^{\lambda_2\lambda}_{\mathbf{q}_{\parallel},\mathbf{p}_{\parallel}} p^{\mu\lambda_2}_{\kp',\kp} - V^{\lambda_2\lambda\mu\lambda_1}_{\kp,\mathbf{q}_{\parallel},\kp',\mathbf{p}_{\parallel}} p^{\lambda\lambda_1}_{\mathbf{q}_{\parallel},\mathbf{p}_{\parallel}} p^{\lambda_1\mu}_{\kp+\qp,\kp'}  \right]  \nonumber \\
&+\sum_{\mu,\lambda,\kp',\mathbf{q}_{\parallel},\mathbf{p}_{\parallel}}\left[  V^{\lambda\mu\lambda_2\lambda_1}_{\mathbf{q}_{\parallel},\kp',\mathbf{p}_{\parallel},\kp+\qp} p^{\lambda\lambda_2}_{\mathbf{q}_{\parallel},\mathbf{p}_{\parallel}} p^{\mu\lambda_2}_{\kp',\kp} - V^{\lambda_2\lambda_1\mu\lambda}_{\kp,\mathbf{q}_{\parallel},\kp',\mathbf{p}_{\parallel}} p^{\lambda_1\lambda}_{\mathbf{q}_{\parallel},\mathbf{p}_{\parallel}} p^{\lambda_1\mu}_{\kp+\qp,\kp'}  \right]  \nonumber \\
&+ i\hbar\partial_t p^{\lambda_1\lambda_2}_{\kp+\qp,\kp} \big|_{\text{coll}} \;. \label{eq:BlochGlRWA}
\end{align}
\end{widetext}
The second line in Eq. \eqref{eq:BlochGlRWA} describes the coupling to the X-ray field. The X-ray transition is initiated by the core occupation $f^{\lambda_1}_{\kp+\qp,\kp-\qp'}$ and blocked by the final band occupation $f^{\lambda_2}_{\kp+\qp+\qp',\kp}$. The third line describes an energy renormalization, due to intraband Coulomb interaction, for both bands. They are well-known in the literature as a core-hole renormalization \cite{Oji1998,Mizoguchi2000,Mauchamp2009} and lead to an effective energetic blue-shift of the transition energy $(E^{\lambda_2}_{\kp}-E^{\lambda_1}_{\kp+\qp})$ of the electronic transition Eq. \eqref{eq:BlochGlRWA} for occupied valence states. The fourth line describes a Coulomb-induced renormalization of the Rabi-frequency $\mathbf{E}_{\qp+\gp}(t)\cdot\mathbf{d}^{\lambda_2\lambda_1}_{\kp,\kp+\qp}(\gp)$, which can be interpreted as local field contribution and leads to the formation of excitons \cite{Bechstedt1980,Olovsson2009,Christiansen2019}. The crystalline excitons are quasi-particles built up by a Coulomb-correlated wave number distribution of core-holes and electrons by two occupied/unoccupied bands, respectively. The equation of motion Eq. \eqref{eq:BlochGlRWA} implicitly includes the Wannier equation \cite{Christiansen2019,Katsch2018} describing the formation of core-excitons bleached by the occupation difference between electrons and core-holes. Depending on the bleaching this contribution to the equation of motion leads to excitonic peaks below the absorption edge. Such bound electron-core-hole quasi-particle can play a considerable role in the interpretation of X-ray spectroscopy \cite{Hjalmarson1981,Olovsson2009,Cocchi2015,Begum2021,Chang2021,Grunes1983}, for example in the case of metal oxides \cite{Grunes1983,Biswas2018,Geneaux2020}. The interpretation of the corresponding two lines can be verified in the limit of a spatially homogeneous system \cite{Kirabuch} and are discussed in the SM Sec. V. An advantageous representation for space-dependent phenomena is the Wigner representation \cite{HessKuhn1996}. The occupations and transitions can be Fourier transformed with respect to their relative momentum. Performing a gradient expansion of the Fourier phase factor and going beyond the zeroth order yields spatially-resolved Bloch equations for occupation and transition in an inhomogeneous system \cite{HessKuhn1996}. The fifth and sixth line include Meitner-Auger-like terms in the Hartree-Fock approximation. As previously discussed, Meitner-Auger interaction couples interband transitions between core and conduction band to transitions within the occupied (or unoccupied) bands. In line five $\mu$ and $\lambda$ need to correspond either both to occupied or unoccupied bands. This leads to a product of two (intersubband) transitions between different conduction or valence bands. Therefore, line five gives rise to nonlinearities in second order. For the first term of the sixth line the band indices have to correspond to two occupied bands, while for the second both need to be unoccupied bands. Also here, the source is quadratic in intersubband transitions and therefore goes beyond the linear optics limit. Within a rotating wave approximation only sources on the right hand side of Eq. \eqref{eq:BlochGlRWA} contribute, which oscillation energy matches the energy difference of $\lambda_1$ and $\lambda_2$ of the left hand side of Eq. \eqref{eq:BlochGlRWA}. Line five and six include the generation of second harmonics resulting from the excitation with intense fields\cite{Garnik2022}.

So far, for the Coulomb contribution no assumption concerning the momentum selection rules were made. Making use of the momentum conservation law $V^{\mu\lambda\nu\rho}_{\mathbf{k}_{\parallel},\mathbf{q}_{\parallel},\mathbf{p}_{\parallel},\mathbf{k}'_{\parallel}}=\sum_{\mathbf{q}'_{\parallel},\gp,\gp'}V^{\mu\lambda\nu\rho}_{\mathbf{k}_{\parallel},\mathbf{q}_{\parallel},\mathbf{p}_{\parallel},\mathbf{k}'_{\parallel}} \delta_{\gp,\kp'-\kp+\mathbf{q}'_{\parallel}}\delta_{\gp',\mathbf{p}_{\parallel}-\mathbf{q}_{\parallel}-\mathbf{q}'_{\parallel}}$ the last four lines (see SM Sec. VI for a detailed derivation) can be written in a more convenient form. Higher-order contributions to the Coulomb interaction, which are included in the collision term, need to be treated on the same level. 
The contribution $\partial_t p^{\lambda_1\lambda_2}_{\kp+\qp,\kp} \big|_{\text{coll}}$ describes the many-particle scattering beyond the Hartree-Fock interaction. The collision term contributes to diagonal $\gamma_{\kp+\qp,\kp}$ and off-diagonal $\mathcal{U}^{\lambda_1\lambda_2}_{\kp+\qp,\kp}$ dephasing\cite{Rossi2002} of the microscopic transition and acts as a dephasing:
\begin{align}
\frac{d}{dt}p^{\lambda_1\lambda_2}_{\kp+\qp,\kp}|_{\text{coll}} &= -\gamma_{\kp+\qp,\kp} p^{\lambda_1\lambda_2}_{\kp+\qp,\kp} + \mathcal{U}^{\lambda_1\lambda_2}_{\kp+\qp,\kp} \;. \label{eq:p_coll}
\end{align}
The diagonal part is determined by the time- and momentum-dependent Coulomb scattering rates
\begin{align}
\gamma_{\kp+\qp,\kp}(t)&=\Gamma^{\text{in}}_{\lambda_1,\kp+\qp}+\Gamma^{\text{out}}_{\lambda_1,\kp+\qp} +\Gamma^{\text{in}}_{\lambda_2,\kp}+\Gamma^{\text{out}}_{\lambda_2,\kp} . \label{eq:gamma}
\end{align}
The off-diagonal contribution couples to all coherences in the Brillouin zone and reads
\begin{align}
\mathcal{U}_{\kp+\qp,\kp}^{\lambda_1\lambda_2}(t)&=\sum_{\mathbf{q},\mathbf{p}} \left(\mathcal{V}^{\lambda_1\lambda_2}_{\kp+\qp,\kp,\mathbf{q},\mathbf{p}} p^{\lambda_1\lambda_2}_{\mathbf{q},\mathbf{p}} + \text{c.c} \right) . \label{eq:offgamma}
\end{align}
For the carrier relaxation processes the Coulomb interaction is treated up to second order Born-Markov approximation. The scattering rates in Eq. \eqref{eq:gamma} and $\mathcal{V}^{\lambda_1\lambda_2}_{\mathbf{k}_1,\mathbf{k}_2,\mathbf{q},\mathbf{p}}$ are specified in the SM Sec. VII. The efficiency of the scattering channels is determined by the Coulomb matrix element and the occupation probabilities of the involved states. The scattering rates feature a sum over different band indices, which include all possible Meitner-Auger relaxation channels, which fulfill momentum and energy conservation at the same time \cite{Winzer2010,Malic2011}. The Meitner-Auger effect \cite{Meitner1922,Auger1923,Grant2004,Matsakis2019} as a non-radiative relaxation mechanism of the core-hole is characterized by the filling of the inner-shell vacancy accompanied by the emission of an electron into the unoccupied conduction band states and possibly even out of the sample into the ionization continuum.

Finally, we present the dynamics in a many band system of the occupations in band $\lambda_1$:
\begin{widetext}
\begin{align}
\frac{d}{dt}f^{\lambda_1}_{\kp+\qp,\kp} &= -\frac{1}{i\hbar}\sum_{\lambda,\qp',\gp} \mathbf{E}_{-\qp'+\gp}(t)\cdot\left( \mathbf{X}^{\lambda_1\lambda}_{\kp,\kp-\qp'}(\gp) p^{\lambda_1\lambda}_{\kp+\qp,\kp-\qp'} - \mathbf{X}^{\lambda\lambda_1}_{\kp+\qp+\qp',\kp+\qp}(\gp) p^{\lambda\lambda_1}_{\kp+\qp+\qp',\kp} \right) \nonumber \\
&-\frac{2}{\hbar}\sum_{\lambda,\kp',\mathbf{q}_{\parallel},\mathbf{p}_{\parallel}} \im\left(\left[ V^{\lambda_1\lambda\lambda\lambda_1}_{\mathbf{q}_{\parallel},\kp',\mathbf{p}_{\parallel},\kp+\qp} p^{\lambda\lambda_1}_{\kp',\kp} - V^{\lambda_1\lambda_1\lambda\lambda}_{\kp,\mathbf{q}_{\parallel},\kp',\mathbf{p}_{\parallel}}p^{\lambda_1\lambda}_{\kp+\qp,\kp'}\right]  p^{\lambda_1\lambda}_{\mathbf{q}_{\parallel},\mathbf{p}_{\parallel}}  \right) \nonumber \\
&+\sum_{\substack{\lambda,\mu,\nu \\ \kp',\mathbf{q}_{\parallel},\mathbf{p}_{\parallel}}} \left[ V^{\lambda\mu\nu\lambda_1}_{\mathbf{q}_{\parallel},\kp',\mathbf{p}_{\parallel},\kp+\qp} \sigma^{\lambda\nu}_{\mathbf{q}_{\parallel},\mathbf{p}_{\parallel}} \sigma^{\mu\lambda_2}_{\kp',\kp} - V^{\lambda_1\lambda\mu\nu}_{\kp+\qp,\mathbf{q}_{\parallel},\kp',\mathbf{p}_{\parallel}} \sigma^{\lambda\nu}_{\mathbf{q}_{\parallel},\mathbf{p}_{\parallel}} \sigma^{\lambda_1\mu}_{\kp+\qp,\kp'} \right]  \nonumber \\
&+\sum_{\kp'}\left[\Gamma^{\text{in}}_{\lambda_1,\kp+\qp,\kp'} \left( \delta_{\kp+\qp,\kp'}-f^{\lambda_1}_{\kp',\kp+\qp} \right) +\Gamma^{\text{in}}_{\lambda_1,\kp,\kp'} \left( \delta_{\kp,\kp'}-f^{\lambda_1}_{\kp,\kp'}\right) \right]  \nonumber \\
&-\sum_{\kp'}\left[ \Gamma^{\text{out}}_{\lambda_1,\kp+\qp,\kp'} f^{\lambda_1}_{\kp',\kp+\qp} + \Gamma^{\text{out}}_{\lambda_1,\kp,\kp'} f^{\lambda_1}_{\kp,\kp'} \right]  \;. \label{eq:Dichte}
\end{align}
\end{widetext}
The first line describes the excitation of a non-equilibrium electron distribution in the band $\lambda_1$. The second line describes the nonlinear Coulomb sources of the carrier population. The third line includes Meitner-Auger-type terms. Therefore, the band indices in the Coulomb element need to correspond to three occupied (unoccupied) and one unoccupied (occupied) bands. Then line three carries a product of occupation and interband transition. Note that, again the momentum conservation for the Coulomb interaction can be inserted. The last two lines expresse the Coulomb interaction described by a microscopic Boltzmann-like scattering equation. The scattering rates explicitly include Pauli-blocking terms and are explicitly given in the SM.

Investigating Eq. \eqref{eq:BlochGlRWA} we see that all quantities, except for the electric field, are known. To determine the electric field as an observable self-consistently, we have to solve the wave equation, which is performed in the following section.

\section{Observables and solution of the wave equation} \label{sec:Absorption}
\begin{figure}[t]
\begin{center}
\includegraphics[width=\linewidth]{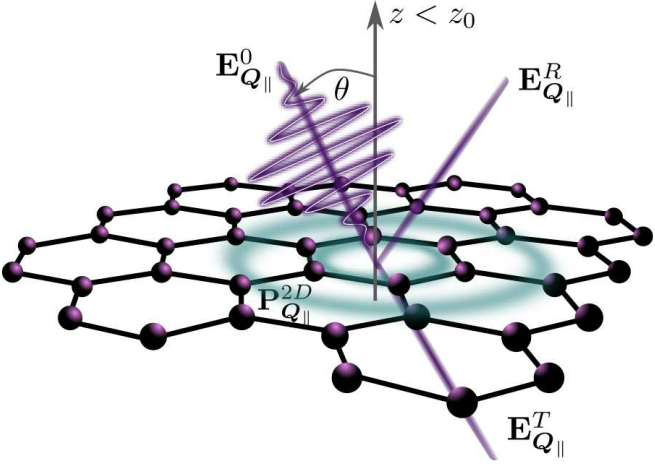}
\end{center}
\caption{Atomically thin material at $z=z_0$ under X-ray radiation $\mathbf{E}^0_{\boldsymbol{Q}_{\parallel}}(z,\omega)$ and two-dimensional response $P^{2D}_{\boldsymbol{Q}_{\parallel}}$ of the material. An incident angle of $\theta=0^{\circ}$ denotes a perpendicular irradiation, while $\theta=90^{\circ}$ corresponds to an in-plane propagating field. $\mathbf{E}^R_{\boldsymbol{Q}_{\parallel}}(z,\omega)$ and $\mathbf{E}^T_{\boldsymbol{Q}_{\parallel}}(z,\omega)$ denote the reflected and transmitted X-ray field, respectively.}
\label{fig:layer}
\end{figure}
Figure \ref{fig:layer} sketches the investigated geometry. A freestanding atomically thin material is irradiated by an external X-ray field $\mathbf{E}^0_{\boldsymbol{Q}_{\parallel}}(z,\omega)$ under an angle of incidence $\theta$. We assume an infinitely thin two-dimensional material located at $z_0$ and use a two-dimensional polarization $\mathbf{P}^{2D}(\mathbf{r}_{\parallel},t)$ to describe the material response $\mathbf{P}(\mathbf{r},t)=\mathbf{P}^{2D}({\mathbf{r}_{\parallel}},t)\delta(z-z_0)$. The measurable observables are the reflected $\mathbf{E}^R_{\boldsymbol{Q}_{\parallel}}(z,\omega)$ and transmitted $\mathbf{E}^T_{\boldsymbol{Q}_{\parallel}}(z,\omega)$ fields containing information on the X-ray material response.

To determine the observables, we solve the Maxwell equations. In the absence of free charges or currents and a spatially homogeneous dielectric environment the corresponding wave equation reads
\begin{align}
\nabla^2\mathbf{E}(\mathbf{r},t)-\frac{\epsilon}{c^2}\frac{\partial^2}{\partial t^2}\mathbf{E}(\mathbf{r},t)&=\mu_0\frac{\partial^2}{\partial t^2}\mathbf{P}(\mathbf{r},t) \nonumber \\
&-\frac{1}{\epsilon_0\epsilon}\nabla\left(\nabla\cdot\mathbf{P}(\mathbf{r},t)\right) \;. \label{eq:waveEq}
\end{align}
Here, for simplicity, we used a uniform background with constant permittivity $\epsilon$ representing off resonant electronic transitions and possible substrate effects. For a free standing layer the dielectric constant reads $\epsilon=1$. Clearly, $\mathbf{P}^{2D}({\mathbf{r}_{\parallel}},t)$ acts as a link between the microscopic X-ray Bloch (cf. Eq. \eqref{eq:obs}) and the wave equations Eq. \eqref{eq:waveEq}. To study experimental observables, we need to solve the Bloch equation Eq. \eqref{eq:BlochGlRWA} and wave equation Eq. \eqref{eq:waveEq} self-consistently.

A formal solution of the wave equation Eq. \eqref{eq:waveEq} is given by the Green's function. Transforming into momentum and frequency space $\mathbf{E}_{\boldsymbol{Q}_{\parallel}}(z,\omega)=\int dz' ~ \mathcal{G}_{\boldsymbol{Q}_{\parallel}}(z-z',\omega)\boldsymbol{\Pi}_{\boldsymbol{Q}_{\parallel}}(z',\omega)$ yields an expression for the in-plane and out-of-plane electric field components as a function of the in-plane radiation wave vector $\boldsymbol{Q}_{\parallel}$, spatial $z$-component and frequency $\omega$ \cite{Mills1975,Knorr1999}. $\boldsymbol{\Pi}_{\boldsymbol{Q}_{\parallel}}(z,\omega)$ denotes the Fourier transform of the source acting on the right hand side in Eq. \eqref{eq:waveEq}. The corresponding Green's function can be determined to be $\mathcal{G}_{\boldsymbol{Q}_{\parallel}}(z-z',\omega)=i\exp(-i\kappa |z-z'|)/2\kappa$ with $\kappa^2=\epsilon\omega^2/c^2-|\boldsymbol{Q}_{\parallel}|^2$. The electric field, separated in in-plane and out-of-plane components and X-ray wave vector restricted to the first Brillouin zone, reads
\begin{widetext}
\begin{align}
\mathbf{E}^{\parallel}_{\qp+\gp}(z,\omega)&= \frac{i}{2\kappa} \left(\frac{(\qp+\gp)\otimes(\qp+\gp)}{\epsilon\epsilon_0} -\mu_0\omega^2\right) \mathbf{P}^{2D}_{\parallel\qp+\gp}(\omega) e^{-i\kappa |z-z_0|} \nonumber \\
&-\frac{i(\qp+\gp)}{2\epsilon\epsilon_0} \text{sgn}(z-z_0) e^{-i\kappa |z-z_0|} P^{2D}_{\perp\qp+\gp}(\omega) \label{eq:Eparallel} \\
E^{\perp}_{\qp+\gp}(z,\omega)&=\frac{i}{2}\left( \frac{\kappa}{\epsilon\epsilon_0}-\frac{\mu_0\omega^2}{\kappa} \right) e^{-i\kappa |z-z_0|} P^{2D}_{\perp\qp+\gp}(\omega) -\frac{i}{2\epsilon\epsilon_0} \text{sgn}(z-z_0) e^{-i\kappa |z-z_0|} (\qp+\gp)\cdot\mathbf{P}^{2D}_{\parallel\qp+\gp}(\omega) \label{eq:Esenk}.
\end{align}
\end{widetext}
with the dyadic product of the X-ray wave vectors $(\qp+\gp)\otimes(\qp+\gp)$. To obtain the full field acting in the Bloch equations Eq. \eqref{eq:BlochGlRWA}, the incident electric field $\mathbf{E}^0_{\mathbf{Q}_{\parallel}+\gp}(z,\omega)$ has to be added to Eq. \eqref{eq:Eparallel} and Eq. \eqref{eq:Esenk} as homogeneous solution. In the limit of $\qp\rightarrow 0$, we find the result of a perpendicular incidence of an optical field on a material sheet, where the field travels as a plane wave varying only in the propagation direction \cite{Graphenbuch}

To describe reflection and transmission, we are interested in the left propagating field in front of the material and the propagating field behind the sample, respectively. Investigating Fig. \ref{fig:layer}, for the reflected light for all $z$ holds $z<z_0$. Therefore, the sign function is negative and the phase factor reads $\exp(i\kappa (z-z_0))$. The reflection coefficient is defined as the ratio of the intensity of the left propagating electric field in front of the material and incident electric field intensity: $r_{\qp}(\omega)=|\mathbf{E}^L_{\qp}(z<z_0,\omega)|^2/|\mathbf{E}_{\qp}^0(z,\omega)|^2$. The transmitted field is located behind the sample, that the sign function becomes positive and propagates in the direction $z>0$, that the phase factor is $\exp(-i\kappa(z-z_0))$. The transmission coefficient is defined as $t_{\qp}(\omega)=|\mathbf{E}^R_{\qp}(z>z_0,\omega)|^2/|\mathbf{E}_{\qp}^0(z,\omega)|^2$ using the right propagating electric field behind the sample \cite{Knorr1996,Katsch2020}. 

Note that, we exploited the two-dimensional nature of the material to access an analytical solution of Maxwell's equation for the propagating X-rays. However, an expansion to three-dimensional bulk materials is possible: In order to expand the theory to three dimensions, the observable Eq. \eqref{eq:obs} needs to be defined in three dimensions including the crystal symmetry, i.e. $\mathbf{P}_{\mathbf{Q}}(t)$. Subsequently, to define observables the Maxwell equations need to be solved by applying additional boundary conditions \cite{Ginzburgbuch,Pollard2009,LaBalle2019}. Here, X-ray propagation effects arise due to the finite sample thickness and the interface needs to be treated carefully, tractable at least numerically. The occurring propagation effects come from interactions between electronic resonances mediated by reemitted photons in an optically thick sample. Propagation induced effects in optically thick samples could be a pulse breakup or polariton beating in case of strong light-matter coupling \cite{Aaviksoo1991,Frohlich1991,Mishina1993}. When the electronic excitations decay radiatively the reemitted photon can propagate giving rise to partial reflections \cite{Manzke1988,Stroucken1995}.

\subsection{Linear X-ray absorption}
We rewrite the electric field as $\mathbf{E}_{\qp+\gp}(z,\omega)=[K_{\qp+\gp}](z,\omega)\mathbf{P}^{2D}_{\qp+\gp}(\omega)+\mathbf{E}^0_{\mathbf{Q}_{\parallel}+\gp}(z,\omega)$ with the matrix
\begin{widetext}
\begin{align}
[K_{\mathbf{Q}_{\parallel}+\gp}](z,\omega) &= \frac{e^{-i\kappa |z-z_0|}}{2i} 
\begin{pmatrix}
\frac{\mu_0\omega^2}{\kappa}-\frac{(Q_x+G_x)^2}{\epsilon_0\epsilon\kappa} & -\frac{(Q_x+G_x)(Q_y+G_y)}{\epsilon_0\epsilon\kappa} & \frac{Q_x+G_x}{\epsilon_0\epsilon}\text{sgn}(z-z_0) \\
-\frac{(Q_y+G_y)(Q_x+G_x)}{\epsilon_0\epsilon\kappa} & \frac{\mu_0\omega^2}{\kappa}-\frac{(Q_y+G_y)^2}{\epsilon_0\epsilon\kappa} & \frac{Q_y+Gy}{\epsilon_0\epsilon}\text{sgn}(z-z_0) \\
\frac{Q_x+G_x}{\epsilon_0\epsilon}\text{sgn}(z-z_0) & \frac{Q_y+G_y}{\epsilon_0\epsilon}\text{sgn}(z-z_0) & \frac{\mu_0\omega^2}{\kappa}-\frac{\kappa}{\epsilon_0\epsilon}
\end{pmatrix} \;.
\end{align}
\end{widetext}
Together with the definition of the macroscopic polarization Eq. \eqref{eq:obs} and the solution of the X-ray Bloch equations in frequency space yields the identification of the susceptibility $[\chi_{\qp}](\omega)$.
Here, in anisotropic media the susceptibility is a second rank tensor since polarization and electric field may not be necessarily collinear anymore. Using the electric field Eq. \eqref{eq:Eparallel} and \eqref{eq:Esenk} at the position $z_0$ and applying the definition for $\mathbf{P}^{2D}_{\mathbf{Q}_{\parallel}+\gp}(\omega)$ yields a self-consistent description of the electric field of the form
\begin{widetext}
\begin{align}
\mathbf{E}_{\mathbf{Q}_{\parallel}+\gp}(z,\omega)&=\epsilon_0[K_{\mathbf{Q}_{\parallel}+\gp}](z,\omega)[\chi_{\mathbf{Q}_{\parallel}}](\omega) \left[\one - \epsilon_0\sum_{\gp'} [K_{\mathbf{Q}_{\parallel}+\gp'}](z_0,\omega)[\chi_{\mathbf{Q}_{\parallel}}](\omega)\right]^{-1}\sum_{\gp''}\mathbf{E}^0_{\mathbf{Q}_{\parallel}+\gp''}(z_0,\omega) + \mathbf{E}^0_{\mathbf{Q}_{\parallel}}(z,\omega) \;. \label{eq:Efield}
\end{align}
\end{widetext}
Equation \eqref{eq:Efield} describes the formation of the reflected and transmitted field as driven by the incident electric field $\mathbf{E}^0_{\mathbf{Q}_{\parallel}+\gp}(z,\omega)$. The occurring dielectric susceptibility is determined by the microscopic Bloch equations, calculated in Sec. \eqref{sec:Bloch}.
For the reflected and transmitted light the sign function and the phase factor in $[K_{\qp+\gp}]$ has to be selected as discussed at the end of the last section. We stress that the relation between X-ray wave vector and angle of incidence is $Q_{\parallel}^2=\omega^2\sin^2\theta/c^2$ obtained from the linear light dispersion.

A complementary approach to include propagation effects is by inserting the field Eq. \eqref{eq:Eparallel} and \eqref{eq:Esenk} in the Bloch equations Eq. \eqref{eq:BlochGlRWA}. Here, besides Coulomb interaction as recombination channel included in the X-ray Bloch equations, a second mechanism of dephasing is caused by radiative interaction: Conduction band electron and the core-hole recombine under the emission of an X-ray photon. This process is included in the X-ray Bloch equation Eq. \eqref{eq:BlochGlRWA} via the self-consistently determined electric field. This allows for a self-consistent description of the radiative dephasing in two-dimensional materials as a function of the wave vector. Inserting the emitted electric field Eq. \eqref{eq:Eparallel} and Eq. \eqref{eq:Esenk} into Eq. \eqref{eq:BlochGlRWA} yields for the radiative contribution (for simplicity here given without the incident field):
\begin{widetext}
\begin{align}
\hbar\omega p^{\lambda_1\lambda_2}_{\kp+\qp,\kp}(\omega)\big|_{\text{rad}}&= \frac{-i}{2\epsilon\epsilon_0}\sum_{\kp',\gp} \left[ \mathbf{d}^{\parallel\lambda_2\lambda_1}_{\kp,\kp+\qp}(\gp) \left(\frac{1}{\kappa}\left((\qp+\gp)\otimes(\qp+\gp)-\epsilon\frac{\omega^2}{c^2}\right)\mathbf{d}^{\parallel\lambda_1\lambda_2}_{\kp'+\qp,\kp'}(\gp) -(\qp+\gp) \right. \right.\nonumber \\
&\hspace{-30mm}\left.\left.\times d^{\perp\lambda_1\lambda_2}_{\kp'+\qp,\kp'}(\gp) \right)+ d^{\perp\lambda_2\lambda_1}_{\kp,\kp+\qp}(\gp)\left( \left(\kappa-\frac{\epsilon\omega^2}{\kappa c^2}\right)d^{\perp}_{\kp'+\qp,\kp'}(\gp)-(\qp+\gp)\cdot\mathbf{d}^{\parallel\lambda_1\lambda_2}_{\kp'+\qp,\kp'}(\gp)\right) \right]p^{\lambda_1\lambda_2}_{\kp'+\qp,\kp'}(\omega) \;. \label{eq:gammaradExakt}
\end{align}
\end{widetext} 
The radiative interaction couples all X-ray induced transitions $p^{\lambda_1\lambda_2}_{\kp+\qp,\kp}$ to all others. This way, the Bloch equation fully include the self-consistent light-matter interaction. To gain more insights in Eq. \eqref{eq:gammaradExakt}, we determine the main contribution to the dephasing from the diagonal dephasing $\kp'=\kp$. The diagonal radiative dephasing reads
 \begin{align}
\gamma^{\text{rad}}_{\kp,\qp}&=\frac{1}{2\epsilon\epsilon_0}\sum_{\gp} \left[  - \epsilon\frac{\omega^2}{c^2}|\mathbf{d}^{\parallel\lambda_1\lambda_2}_{\kp+\qp,\kp}(\gp)|^2 \right.\nonumber \\
&\left.\hspace{-11mm}+\frac{1}{\kappa} \mathbf{d}^{\parallel\lambda_2\lambda_1}_{\kp,\kp+\qp}(\gp)(\qp+\gp)\otimes(\qp+\gp)\mathbf{d}^{\parallel\lambda_1\lambda_2}_{\kp+\qp,\kp}(\gp) \right.\nonumber \\
&\left.\hspace{-11mm}+(\kappa-\frac{\epsilon\omega^2}{\kappa c^2})|d^{\perp\lambda_1\lambda_2}_{\kp+\qp,\kp}(\gp)|^2 \right. \nonumber \\
&\left.\hspace{-11mm}- 2\re((\qp+\gp)\cdot\mathbf{d}^{\parallel\lambda_1\lambda_2}_{\kp+\qp,\kp}(\gp) d^{\perp\lambda_1\lambda_2}_{\kp,\kp+\qp}(\gp))\right] \;. \label{eq:gammarad}
\end{align} 
For a vanishing $\qp+\gp$ (normal incidence) only the first term of Eq. \eqref{eq:gammarad} survives. In accordance to a first order perturbation theory, Eq. \eqref{eq:gammarad} resembles Fermi's golden rule for the radiative broadening and energy shifts, both, determined by the square of the dipole matrix element.

To compare our result with the literature, the rate of spontaneous emission $W_X$ between an initial band $\lambda$ and the 1$s$-band for atoms is calculated in dipole approximation by Fermi's golden rule \cite{HendersonBuch}:
\begin{align}
W_X\propto\omega^3 \sum_{\beta,j} |\langle 1s,\beta,1s,0 \mid \mathbf{r} \mid \lambda,\beta,j,0 \rangle |^2 
\end{align}
where $\omega$ stands for the transition energy between initial and final state. While we find for the two-dimensional semiconductor a $\omega$-dependence of the dephasing, cf. Eq. \eqref{eq:gammarad} with inserted $\qp$, for atoms a $\omega^3$-dependence is well known. The difference originates from the different dimensionality of the emitting systems. To connect our results to the existing literature we specify to the atomic case: Here, we can argue that the electronic wave functions have strongly contributing values only for $|\mathbf{r}|<a_B/Z$ and we can roughly approximate the transition integral to be proportional to $Z^{-1}$. From Moseley's law \cite{Moseley1913}, being an extension of the Rydberg formula, we know that the transition energy $\omega$ is proportional to $Z^2$. Consequently, we see that the spontaneous K-shell emission rate is proportional to $Z^4$ ~ \cite{KuzmanyBuch}. It can be shown that for atoms the Meitner-Auger yield $W_A$ is almost independent of the effective nuclear charge \cite{FeldmannBuch}. A semi-empirical expression for the X-ray yield is introduced in Ref. \cite{Fink1966}:
\begin{align}
w_X=\frac{W_X}{W_X+W_A} \;,
\end{align}
which is proportional to $Z^4$. Hence, for atomic systems Meitner-Auger transitions offer higher sensitivity to core level transitions for low $Z$ and luminescence is more likely to occur with increasing nuclear charge. The advantage of the formalism presented here is that for radiative and Meitner-Auger recombination, it might be possible to extent and establish similar statements for crystalline solids. For example, the luminescence yield can be calculated from the time-integrated conduction band electron occupation, which decays radiatively $W_X=\sum_{\kp,\qp}\gamma^{\text{rad}}_{\kp,\qp} f^c_{\kp}$. In a similar way also the amount of non-radiatively decaying electrons can be calculated from the microscopic scattering rates expressed by the X-ray Bloch equations. 

\section{Application to graphene} \label{sec:Graphene}
We apply the derived theory to the exemplary material graphene \cite{Graphenbuch,Reichbuch}. 
Graphene is a monolayer of carbon atoms in a two-dimensional honeycomb lattice. Carbon exhibits six electrons with an electronic ground state configuration of 1$s^2$2$s^2$2$p^2$. The 1$s$-electrons form the core bands, the 2$s$ and 2$p$ electrons are valence electrons and dominate the energy dispersion around the Fermi level, which is set to \unit[0]{eV} for undoped graphene. The 2$s$ and 2$p_{x/y}$-electrons undergo a sp$^2$-hybridization with one further electron in the 2$p_z$-orbital. The in-plane hybridized electrons form the $\sigma$-bands, which are responsible for the in-plane covalent bonding between the atoms. The 2$p_z$-electrons form the $\pi$-bands \cite{Dresselhausbuch}.

\begin{figure}[t]
\centering
\includegraphics[width=\linewidth]{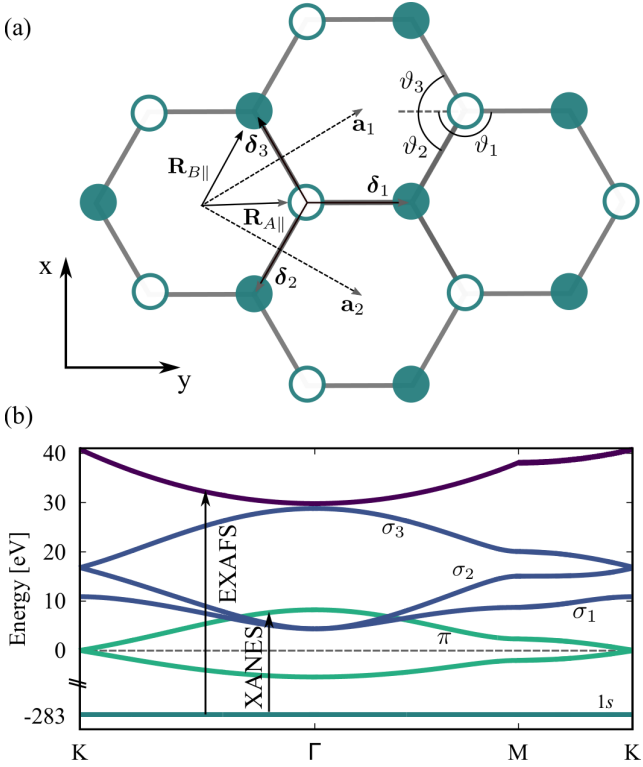}
\caption{(a) Hexagonal lattice of graphene. Empty circles denote the atoms on sublattice A and filled circles on sublattice B. The vectors $\boldsymbol{\delta}_i$ denote the nearest-neighbor vectors connecting atom A to the three surrounding B atoms. (b) Electronic dispersion above the Fermi level (grey dashed) and for orientation also the filled $\pi$-band below \unit[0]{eV}. The light green curves show the $\pi$-bands and the blue curves show the $\sigma$-bands. Dark green denotes the degenerate core band and violet the ionization continuum for $k_z=0$.}
\label{fig:lattice}
\end{figure}

Figure \ref{fig:lattice}(a) illustrates the unit cell, spanned by the lattice vectors $\mathbf{a}_1$ and $\mathbf{a}_2$, containing two atoms and constituting the equivalent sublattices A and B, which are rotated with respect to each other by $\pi/3$. The vectors $\mathbf{R}_{A\parallel}$ and $\mathbf{R}_{B\parallel}$ denote the position of the corresponding atoms obtained by linear combination of the lattice vectors. In nearest-neighbor approximation the constructed vectors $\boldsymbol{\delta}_{BA}=\mathbf{R}_{B\parallel i}-\mathbf{R}_{A\parallel}\equiv\boldsymbol{\delta}_i$ connect the atoms on sublattice A with the three surrounding atoms on sublattice B, cf. Fig \ref{fig:lattice}(a). The six second nearest next-neighbors are located on the same sublattice.

\subsection{Band structure and dipole moments of graphene}
The energy dispersion and corresponding tight binding coefficients are obtained from the Schr\"odinger equation Eq. \eqref{eq:SG} \cite{Dresselhausbuch}, cf. Fig. \ref{fig:lattice} (b) which also includes the corresponding XANES and EXAFS transitions. The Hamiltonian of graphene is symmetric with respect to the $x$-$y$ plane. Consequently, the $\pi$ and $\sigma$-bands decouple since the former have an odd symmetry and the latter even with respect to the reflection. For the $\pi$-bands we obtain analytically \cite{Reich2002}
\begin{align}
E_{\pi,\mathbf{k}_{\parallel}}=\pm t_{\pi} |\xi_{\mathbf{k}_{\parallel}}| \label{eq:piband}
\end{align}
with the positive solution describing the valence band and the negative solution the conduction band formed of 2$p_z$-electrons ($t_{\pi}<0$). The shape of the band structure depends on the lattice symmetry only, described by the nearest-neighbor form factor $\xi_{\mathbf{k}_{\parallel}}=\sum_{i}\exp(i\mathbf{k}_{\parallel}\cdot\boldsymbol{\delta}_i)$. Here, for the analytical treatment we neglected contributions from the overlap of electronic orbitals, which is generally small and lead to an asymmetry between the valence and conduction $\pi$-band. Since we include the full form factor $\xi_{\kp}$ without Taylor expansion trigonal warping is included \cite{Rakyta2010,Rose2013}. The electronic hopping between 2$p_z$-orbitals from different lattice sites is denoted by $t_{\pi}$. The band exhibits Dirac cone-like minima at the K-points, saddle points at the M-points and a maximum at the $\Gamma$-point, all included in Eq. \eqref{eq:piband}. As discussed in Sec. \ref{sec:Hamilton1}, the core dispersion is treated as a flat band $E^i=-\unit[283]{eV}$ \cite{Susi2015} and the ionization continuum as a manifold of three-dimensional parabolas. Because of the neglected hopping for the core states between different sublattices the bonding and anti-bonding core state are degenerate. The tight binding coefficients for $\pi$ and $1s$-state read $C_{A 2p_z/1s,\kp}^{\pi/1s}=\mp\xi_{\kp}/\sqrt{2}|\xi_{\kp}|$ and $C^{\pi/1s}_{B 2p_z/1s,\kp}=1/\sqrt{2}$. In theoretical approaches focusing on the response with respect to optical frequencies in the visible range the orbital composition of graphene is restricted to the 2$p_z$-orbital, which governs the optical properties of graphene. The reason for this approximation that only the K and M-points of the $\pi$-bands lie energetically in the optical range of the Fermi surface. However, in X-ray experiments the excitation energy is tuned over several hundreds of eV. Therefore, also states energetically further away from the Fermi level have to be taken into account. The $\sigma$-bands are obtained by numerical diagonalization of the Schr\"odinger equation. A spaghetti plot of the used band structure is displayed in Fig. \ref{fig:lattice}(b). Note, that for the numerical evaluation we focus on an analytical 10-band tight-binding model as proof of principle and as illustration of the approach. However, to fully describe linear X-ray absorption experiments over several hundreds of eV an accurate band structure beyond five orbitals, as it appears in Eq. \eqref{eq:obs}, from \textit{ab initio} electronic structure theory, as provided by density function theory (DFT) calculations are necessary. In this paper we use the 5-orbital system as a toy model only. In the following, we will always refer with $\pi$ and $\sigma$ to the unoccupied conduction band states.

With the tight-binding coefficients $C^{\lambda}_{j\beta,\kp}$ we can calculate the XANES dipole matrix element Eq. \eqref{eq:XANEStb} for the core-conduction band transitions. Since we are interested in excited interband transitions the intraband contribution is neglected. Restricting the next-neighbor sum to the first nearest-neighbors we obtain in dipole approximation for the K-edge dipole matrix elements
\begin{align}
\mathbf{X}^{1s\lambda}_{\mathbf{k}_{\parallel}+\mathbf{Q}_{\parallel},\mathbf{k}_{\parallel}}&=-e_0\sum_{\alpha,\beta,i} C^{*1s}_{\beta 1s,\mathbf{k_{\parallel}+Q_{\parallel}}}C^{\lambda}_{\alpha i,\mathbf{k}_{\parallel}} \nonumber \\
&\times \left(\langle 1s,\beta,1s,0 \mid \mathbf{r}  \mid \lambda,\alpha,i,0\rangle ~ \delta_{\alpha,\beta} \right. \nonumber \\
&\left. + e^{i\mathbf{k_{\parallel}}\cdot\boldsymbol{\delta}_{\beta\alpha}} \langle 1s,\beta,1s,0 \mid \mathbf{r}  \mid \lambda,\alpha,i,\boldsymbol{\delta}_{\beta\alpha}\rangle \right) \;. \label{eq:X}
\end{align}
The dipole matrix element exhibits an onsite (first term) and an offsite contribution (second term). The band $\lambda=\pi$ consists just of the $i=2p_z$ orbital. The $\lambda=\sigma$-band is built by $i=2s,2p_x,2p_y$ orbitals. The offsite transition integral between sublattices A and B can be calculated analytically by transforming the integral to prolate spheroidal coordinates helping to handle the two-center nature of the integral. The calculation can be found in the SM Sec. VIII. Here, we briefly present the result of the calulations: The $z$-polarized $\pi$-transition depends only on the absolute value of the next-neighbor vector, which is $\delta_i=\unit[0.14]{nm}$ for the nearest-neighbors. The corresponding integral has a value of
\begin{align}
\langle 1s,A,1s,0\mid\mathbf{r}\mid\pi,B,2p_z,0.14\rangle = \begin{pmatrix}
0 \\
0 \\
0.14
\end{pmatrix} \text{pm} \;.
\end{align}
In contrast the transition into the $\sigma$-bands are in-plane polarized because of the mirror symmetry of the graphene plane. The in-plane transitions are differently weighted for each next-neighbor vector depending on its angular orientation $\vartheta_i$ to the $y$-axis, cf. Fig. \ref{fig:lattice}(a). We find
\begin{align}
\langle 1s,A,1s,0\mid\mathbf{r}\mid\sigma,B,2p_x,\boldsymbol{\delta}_i\rangle=\nonumber \\
&\hspace{-40mm}\begin{pmatrix}
\cos^2\vartheta_i & \sin^2\vartheta_i & 0 \\
\sin\vartheta_i\cos\vartheta_i & -\sin\vartheta_i\cos\vartheta_i & 0 \\
0 & 0 & 1
\end{pmatrix}
\begin{pmatrix}
-0.12 \\
0.14 \\
0
\end{pmatrix} \text{pm}
\nonumber \\
\langle 1s,A,1s,0\mid\mathbf{r}\mid\sigma,B,2p_y,\boldsymbol{\delta}_i\rangle=\nonumber \\
&\hspace{-40mm}\begin{pmatrix}
\sin\vartheta_i\cos\vartheta_i & -\sin\vartheta_i\cos\vartheta_i & 0 \\
\sin^2\vartheta_i &\cos^2\vartheta_i & 0 \\
0 & 0 & 1
\end{pmatrix}
\begin{pmatrix}
-0.12 \\
0.14 \\
0
\end{pmatrix} \text{pm} \;. \label{eq:1s_p}
\end{align}
In the $\delta\rightarrow 0$ limit the prolate spheroidal coordinates evolve into the spherical coordinates. The onsite transitions for the $2p_x$, $2p_y$ and $2p_z$-integrals are purely $x$, $y$ and $z$-polarized, respectively, and have the value of \unit[4.00]{pm}. Finally, the 1$s$ to 2$s$ transition does not contribute for vanishing connection vector because of there even symmetry and the uneven parity of the dipole vector. For the offsite transition we obtain
\begin{align}
\langle 1s,A,1s,0\mid\mathbf{r}\mid\sigma,B,2s,\boldsymbol{\delta}_i\rangle=-0.5
\begin{pmatrix}
\cos\vartheta_i \\
\sin\vartheta_i \\
0
\end{pmatrix} \text{pm} \;.
\end{align}
The result is negative because of the node of the $2s$-orbital. Obviously, offsite transitions $(\delta\neq 0)$ are smaller than onsite transition $(\delta=0)$ integrals because of the small overlap of the core orbital with other orbitals. The transition integral of the EXAFS matrix element for 1$s$-electrons has already been calculated in Eq. \eqref{eq:1sexafs}.

Next, we evaluate the X-ray Bloch equations. For a weak X-ray excitation density we assume that the occupations are unchanged by the optics corresponding spatially homogeneous system ($f^{\lambda}_{\mathbf{k}_1,\mathbf{k}_2}\rightarrow f^{\lambda}_{\mathbf{k}_1,\mathbf{k}_2}\delta_{\mathbf{k}_1,\mathbf{k}_2}$) and set the diagonal occupation in initial and final band to one and zero, respectively. Then, in linear optics, the solution of the microscopic Bloch equation Eq. \eqref{eq:BlochGlRWA} reads, with dipole approximated matrix element, in frequency space
\begin{align}
p^{\lambda_1\lambda_2}_{\mathbf{k}_{\parallel}+\mathbf{Q}_{\parallel},\kp}(\omega)&=\frac{-\sum_{\gp}\mathbf{X}^{\lambda_2\lambda_1}_{\mathbf{k}_{\parallel}+\mathbf{Q}_{\parallel},\kp}\cdot \mathbf{E}_{\mathbf{Q}_{\parallel}+\gp}(\omega)}{\hbar\omega-E^{\lambda_2}_{\kp}+E^{\lambda_1}_{\mathbf{k}_{\parallel}+\mathbf{Q}_{\parallel}} + i\gamma } \label{eq:lsg} \\
p^{\lambda k_z}_{\mathbf{k}_{\parallel}+\mathbf{Q}_{\parallel},\kp}(\omega)&=\frac{-\sum_{\gp}\mathbf{Y}^{k_z\lambda}_{\mathbf{k}_{\parallel}+\mathbf{Q}_{\parallel},\kp}\cdot \mathbf{E}_{\mathbf{Q}_{\parallel}+\gp}(\omega)}{\hbar\omega-E^{k_z}_{\kp}+E^{\lambda}_{\mathbf{k}_{\parallel}+\mathbf{Q}_{\parallel}} + i\gamma } \label{eq:lsg2}
\end{align}
for the XANES and EXAFS transitions, respectively. The former equation describes absorption between two-dimensional bands within the material (1$s$ to conduction band) and the latter yields the absorption of 1$s$ to vacuum transitions. Coulomb-induced renormalizations of the unexcited graphene band gap are assumed to be included in the single-particle flat band core binding energy $E_{1s}$ \cite{Susi2015}.As a first attack, and as experiments suggest, core-excitonic effects of graphene are neglected in Eq. \eqref{eq:lsg}. This allows to illustrate the strength of including the Bloch theorem in crystalline solids via a first analytical approach Eq. \eqref{eq:lsg} and \eqref{eq:lsg2}. Additionally, we introduced a phenomenological dephasing constant $\gamma$ attributing to the finite lifetime of the electronic transition and leading to a broadening of the absorption line. 
The radiative lifetime for graphene is of a few picoseconds obtained from Eq. \eqref{eq:gammarad}. Different studies report a core-hole lifetime in graphite and carbon-based molecules of around \unit[10]{fs} \cite{Harada2004,Schlachter2004,Mucke2015}. However, the formalism allows for a straightforward calculation of the linewidth due to Coulomb interaction (cf. Eq. \eqref{eq:p_coll}) or phonon scattering \cite{Selig2016,Christiansen2017,Bernardi2021}. Together with the definition of the two-dimensional macroscopic polarization, being proportional to the electric field, we can identify the dyadic susceptibility
\begin{figure}[t]
\begin{center}
\includegraphics[width=\linewidth]{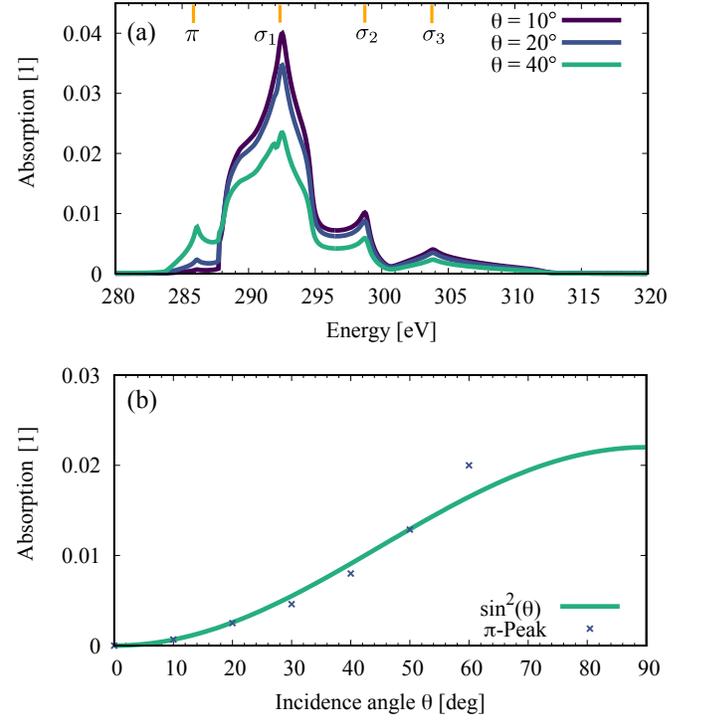}
\end{center}
\caption{(a) XANES of graphene for different angles of incidence. The first peak stems from absorption into the $\pi$-bands, while the three succeeding peaks are transitions into the $\sigma$-bands. (b) The $\pi$-peak as extracted from the numerics shows a $\sin^2\theta$-behavior. The deviation at higher angles stems from the fact that we assumed an infinite thin graphene layer.}
\label{fig:xanes}
\end{figure}
\begin{align}
[\chi_{\mathbf{Q}_{\parallel}}](\omega)&=\sum_{\lambda_1,\lambda_2,\kp}\frac{\mathbf{X}^{\lambda_1\lambda_2}_{\kp,\mathbf{k}_{\parallel}+\mathbf{Q}_{\parallel}}\otimes\mathbf{X}^{\lambda_2\lambda_1}_{\mathbf{k}_{\parallel}+\mathbf{Q}_{\parallel},\kp}}{\hbar\omega-E^{\lambda_2}_{\kp}+E^{\lambda_1}_{\mathbf{k}_{\parallel}+\mathbf{Q}_{\parallel}} + i\gamma} \nonumber \\
&+\sum_{\lambda,\kp,k_z}\frac{\mathbf{Y}^{\lambda k_z}_{\kp,\mathbf{k}_{\parallel}+\mathbf{Q}_{\parallel}}\otimes\mathbf{Y}^{k_z\lambda}_{\mathbf{k}_{\parallel}+\mathbf{Q}_{\parallel},\kp}}{\hbar\omega-E^{k_z}_{\kp}+E^{\lambda}_{\mathbf{k}_{\parallel}+\mathbf{Q}_{\parallel}} + i\gamma} \label{eq:susc}
\end{align}
with the dyadic product of the dipole matrix elements yielding a $3\times 3$-matrix. The XANES dipole matrix element $\mathbf{X}^{\lambda_2\lambda_1}_{\mathbf{k}_{\parallel}+\mathbf{Q}_{\parallel},\kp}$ is determined by Eq. \eqref{eq:X} and the EXAFS matrix element $\mathbf{Y}^{\lambda k_z}_{\kp,\mathbf{k}_{\parallel}+\mathbf{Q}_{\parallel}}$ by Eq. \eqref{eq:1sexafs}. The X-ray wave number and the excitation frequency are related by the relation $Q^2_{\parallel}=\omega^2\sin\theta/c^2$, where $\theta$ denotes the incident angle of the X-ray radiation and $c$ the speed of light in the surrounding medium (cp. Sec. \eqref{sec:Absorption}). Consequently, we perform no further approximation and include all $\qp$ over the full range of the absorption spectrum. With the susceptibility all quantities are known and the response of the material to the weak excitation is investigated by calculating the absorption. Therefore, we introduce with the dipole approximated dipole matrix element the absorption coefficient as $\alpha_{\qp}(\omega)=1-r_{\qp}(\omega)-t_{\qp}(\omega)$ defined by reflection $r_{\qp}$ and transmission $t_{\qp}$. The reflection is defined by the reflected intensity of the left propagating field in front of the graphene sheet $r_{\qp}(\omega)=I^L_{\qp}(\omega)/I^0_{\qp}(\omega)$ and the transmission by the intensity of the right propagating electric field behind the graphene sheet $t_{\qp}(\omega)=I^R_{\qp}(\omega)/I^0_{\qp}(\omega)$.

\subsection{X-ray spectrum of graphene}
Due to the non-isotropic character of graphene, in the susceptibility all possible components of the dipole matrix element are coupled, which leads to a rather difficult expression for the absorption. We start by investigating the XANES spectrum, i.e. the susceptibility in Eq. \eqref{eq:susc} is restricted to the first line and calculating the absorption with Eq.\eqref{eq:Efield}.
\begin{figure}[t]
\begin{center}
\includegraphics[width=\linewidth]{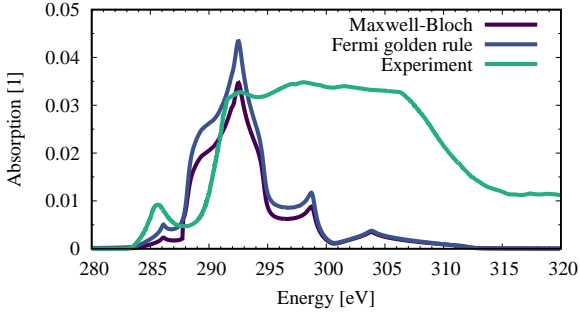}
\end{center}
\caption{Comparison between the true absorption coefficient, Fermi's golden rule and experiment. Due to a weak X-ray-matter interaction both calculated absorption curves are almost identical. They reproduce the first four resonances of the experiment but miss the plateau at higher energies since these electronic states are not included in the numerical evaluation. The experimental curve is adjusted in height to the calculated absorption spectrum.} 
\label{fig:comparison}
\end{figure}
Figure \ref{fig:xanes}(a) displays the calculated XANES spectrum of graphene for different angles of incidence $\theta$. The absorption starts at \unit[283]{eV} corresponding to the binding energy of the core electrons relative to the Fermi surface of graphene. The first peak at \unit[286]{eV} rises with increasing angle and stems from the transition of 1$s$-electrons into the $\pi$-band close to the M-point, which  exhibits a van Hove singularity due to a saddle point. The following three peaks at \unit[293]{eV}, \unit[298]{eV}, and \unit[303]{eV} stem from transitions into the three $\sigma$-bands and are decreasing with angle of incidence. All four peaks are observed in experiments for graphite \cite{Rosenberg1986,Buades2018} and graphene \cite{Pacile2008,Papagno2009,Xu2015} and matches with the here calculated energetic positions and spectral structure. However, when comparing Fig. \ref{fig:xanes}(a) to experiments \cite{Rosenberg1986,Buades2018,Pacile2008,Papagno2009,Xu2015} a clear difference can be noticed (cp. green line in Fig. \ref{fig:comparison}): After the second peak a plateau in the absorption spectrum is observed. This can be traced back to the point that we limited our approach to orbitals up to the 2$p$-orbitals (10 band approximation). Including more orbitals \cite{Willis1974} leads to a continuum of close lying bands and should form an almost continuous absorption. At this point, we want to point out, that the strength of the developed Bloch equation approach lies in the description of self-consistent many-body interaction, as introduced in Sec. \ref{sec:Bloch}, and the description of nonlinear optics occurring in ultrafast pump-probe type experiments as attosecond transient absorption. The use of the tight binding approach to calculate the band structure as input (as done here for illustration) is not always sufficient (illustrated later on in Fig. \ref{fig:comparison}) but a connection of the Maxwell-X-ray Bloch formalism developed here and more advanced DFT methods, as $\Delta$SCF-DFT \cite{Liang2017,Klein2021}, for single-particle energies and coupling elements should be applied to accurately predict X-ray-matter interaction and nonlinear effects. In our first attack here, the tight binding method is just used for illustration.

Figure \ref{fig:xanes}(b) shows the peak height of the $\pi$-peak as a function of the angle of incidence. We obtain that the absorption follows a $\sin^2\theta$-behavior before deviating from this trend at about \unit[50]{$^{\circ}$}: This polarization-dependent absorption can be explained by the symmetry of the orbital composition of the final band as was already done in different theoretical and experimental works \cite{Chowdhury2012,Rosenberg1986,Schiessling2003,HemrajBenny2006}. Because of the out-of-plane character of the 2$p_z$-orbital the dipole transition for the $\pi$-peak is $z$-polarized, which greatly simplify the susceptibility tensor since only one component remains nonzero. Consequently, we can find an analytical expression for the absorption into the $\pi$-band. Without the loss of generality we assume for the in-plane incidence angle zero degree, such that $Q_y=0$. Then we obtain for small $\qp$ for the absorption
\begin{align}
\alpha_{\qp}(\omega)=\frac{\Delta_{\qp}(\omega)\im\left(\chi^{33}_{\qp}(\omega)\right)-\frac{Q_x^2}{2\epsilon}|\chi^{33}_{\qp}(\omega)|^2}{|1-\frac{i}{2}\Delta_{\qp}(\omega)\chi^{33}_{\qp}(\omega)|^2} \label{eq:abs_exact}
\end{align}
with $\Delta_{\qp}(\omega)=\frac{\omega^2}{c^2\kappa} -\frac{\kappa}{\epsilon}$ and $\chi^{33}_{\qp}(\omega)$ as the $zz$-entry of the susceptibility tensor. The coefficient $\Delta_{\qp}(\omega)$ stems from the matrix $[K_{\qp}](\omega)$ following from the Maxwell equations. In particular, the denominator arises from the self-consistent treatment of Maxwell and Bloch equations and includes for instance the radiative coupling in the sample. The in-plane light wave vector $|\qp|^2$, which is orthogonal to the X-ray polarization vector of the incident light, is proportional to $\sin^2\theta$. When inserting the definition for $\kappa=\sqrt{\epsilon\omega^2/c^2-|\qp|^2}$ and express the X-ray wave vector as function of the angle of incidence we obtain for the prefactor $\Delta_{\qp}=\omega^2\sin^2\theta/\epsilon c^2\sqrt{\epsilon-\sin^2\theta}$. Now, we explicitly observe the $\sin^2$-dependence of the true absorption coefficient for the $\pi$-transition. For a perpendicular irradiation of the sample $\theta=0^{\circ}$ we see immediately that $\Delta_{\qp}(\omega)$ and $Q_x$ vanish and consequently also the absorption vanishes as expected from the $z$-polarized transition. The derived absorption formula diverges for an incident angle of $\theta=90^{\circ}$ in the case of $\epsilon=1$, explaining the observed deviation in Fig. \ref{fig:xanes}(b). That for $\epsilon>1$ the singularity does not explicitly appear anymore, suggests that the origin lies in the assumption of an infinitely thin layer in vacuum. The divergence can be lifted by starting with a three layer model -- supstrate, layer, substrate -- all with finite thickness and solve the Maxwell equation for each region with corresponding continuity conditions \cite{Chewbuch}. In our case, because of the weak X-ray matter interaction, the additional denominator in Eq. \eqref{eq:abs_exact} including the radiative coupling plays only a minor role. We elaborate on this in the following.

In Fig. \ref{fig:comparison}, we plot the 20$^{\circ}$-result from Fig. \ref{fig:xanes}(a) and a XANES measurement on graphene taken from Ref. \cite{Papagno2009} (16$^{\circ}$). As discussed previously, we can see that the four resonances, which are explicitly calculated, reproduces well the experiment, but that the plateau is missed due to the fail of the restricted tight binding method (cp. above) applied in our approach. To evaluate the influence of X-ray propagation effects, included in the denominator unequal to one in Eq. \eqref{eq:abs_exact}, we plot the result obtained from Fermi's golden rule, which neglects propagation effects. To obtain Fermi's golden rule, we start from the X-ray Bloch equation Eq. \eqref{eq:lsg} and carry out the limit $\gamma\rightarrow 0$. The intensity of a light field propagating though a medium is damped by the imaginary part of the complex permittivity \cite{Kochbuch}. Therefore, we can approximate the absorption by $\alpha(\omega)=\omega\im(\chi(\omega))/cn=\omega\im(P(\omega)/E(\omega))/\epsilon_0cn$ \cite{Kochbuch}, where we projected the polarization density on the polarization vector and introduced the material refractive index $n$. With the macroscopic polarization density we obtain an expression for the absorption coefficient for mall $\qp$ in terms of Fermi's golden rule
\begin{align}
\alpha_{\qp}(\omega)&=\frac{\omega\pi}{\epsilon_0cn}\frac{A}{4\pi^2}\sum_{\lambda} \int d^2k_{\parallel} ~ |X_{\kp+\qp,\kp}^{1s\lambda}|^2 \nonumber \\
&\times \delta\left(E^{\lambda}_{\kp}-E^{1s}_{\kp+\qp}-\hbar\omega\right) \;, \label{eq:FermiAbs}
\end{align}
which we can fairly compare to the self-consistent result. In Eq. \eqref{eq:FermiAbs} we transformed the $\kp$-sum into an integral with area $A$. To have a fair comparison with our result, the delta function is phenomenologically broadened by the same rate $\gamma$ as the X-ray Bloch equations. Figure \ref{fig:comparison} shows the result of Fermi's golden rule Eq. \eqref{eq:FermiAbs}, which well reproduces the self-consistent result: We see that the magnitude of the results are comparable. The reason lies in the weak X-ray-matter interaction. Differences between Fermi's golden rule and the true absorption can be expected, when non-markovian effects such as quantum kinetics of electron-phonon interactions (if not included in Fermi's golden rule) dominate the lineshape or strong radiative interaction occurs in a way that the perturbation approach of Eq. \eqref{eq:FermiAbs} breaks down. The latter can for example occur for two-dimensional materials involving more layers enhancing the light-matter interaction: For example, taking ten two-dimensional sheets can lead to an increase of the susceptibility by a factor of 10 ~\cite{Graphenbuch}. Increasing the susceptibility leads to a deviation of Fermi's golden rule and the true absorption\cite{Stroucken1996}. In this case, the true absorption becomes suppressed and accumulates an increased radiative broadening, cf. Fig. S2 of the SM. Also the oscillator strength can be distributed differently, due to lineshifts. Other scenarios, where Fermi's golden rule breaks down are X-ray excitations of materials in a cavity or for materials, which have an intrinsically stronger light-matter interaction. For the latter nonlinear many-body effects can show up differently in transmission and reflection geometry \cite{Katsch2020_2} and reflection is not included in Fermi's golden rule.
\begin{figure}[t]
\begin{center}
\includegraphics[width=\linewidth]{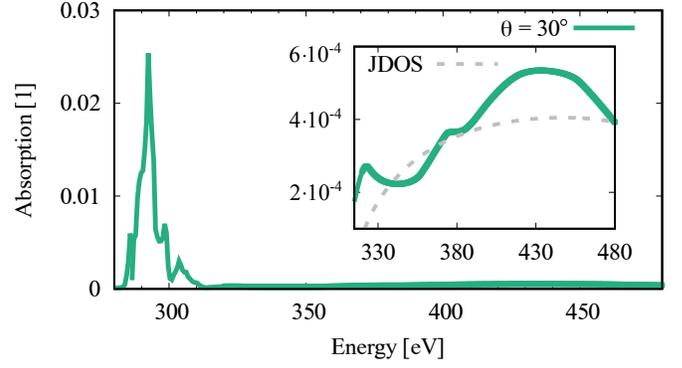}
\end{center}
\caption{X-ray absorption spectrum of graphene for $30^{\circ}$ irradiation. We obtain a dominating XANES contribution stemming from transition between core and conduction bands. Also, the weaker transitions into the ionization continuum, displayed in the inset, are obtained. The absorption into the three-dimensional continuum (grey dashed displays the joint density of states) is modulated by oscillations with maxima at different energies.}
\label{fig:exafs}
\end{figure}

After investigating the XANES contribution, we compute the full X-ray absorption spectrum by including also the EXAFS part, described by the second line of Eq. \eqref{eq:susc}. The ionization threshold is set artificially to the maximum of the energetically highest lying $\sigma$-band. Figure \ref{fig:exafs} shows the full absorption spectrum ($\theta=30^{\circ}$) and for better visibility the EXAFS part as inset, where the absorption involving the three-dimensional electronic continuum of vacuum states as final states can be observed. The absorption curve is modulated with oscillations. For the particular situation plotted here we recognize three maxima. Following the pioneering work by Sayers \textit{et al.} it is widely recognized that these wiggles could be used to obtain quantitative information about the spatial structure of the \blue{atomic} lattice the X-rays are interacting with. The current interpretation in literature is that these oscillations stem from a modulation of the absorption cross section due to interference of the X-ray waves between neighboring atoms. In our approach, which is specifically developed for crystalline solids, includes the full lattice symmetry by the sum over reciprocal lattice vectors, we calculate the absorption using the susceptibility determined by the second line of Eq. \eqref{eq:susc}, which naturally includes the Bloch theorem. For a constant EXAFS dipole matrix element the second line of Eq. \eqref{eq:susc} summed over the electronic wave vectors describes a square-root like absorption into the three-dimensional continuum states of the vacuum electrons, cf. the grey dashed line of Fig. 8 (inset). However, this simplified absorption line is modulated by the dyadic product of the EXAFS matrix elements, cf. Eq. \eqref{eq:susc}. The occurring form factor of the EXAFS matrix element is abbreviated $f_n(\mathbf{k})=8\pi i\zeta^{3/2}\mathbf{k}_n\hat{\mathbf{e}}_n/(\zeta^2+(\mathbf{k})^2)^3$, cf. Eq. \eqref{eq:1sexafs}. Analytically we obtain for the squared dipole matrix element Eq. \eqref{eq:1sexafs} of the entry $nm$ in the susceptibility tensor Eq. \eqref{eq:susc}
\begin{align}
|(C^{*1s}_{A 1s,\mathbf{k}_{\parallel}}+C^{*1s}_{B 1s,\mathbf{k}_{\parallel}})|^2f_n(\mathbf{k}) f_m(\mathbf{k}) = \nonumber \\
\left(1+\frac{1}{|\xi_{\mathbf{k}_{\parallel}}|}\sum_{i}\cos(\mathbf{k}_{\parallel}\cdot\boldsymbol{\delta}_i)\right)f_n(\mathbf{k}) f_m(\mathbf{k})  \label{eq:anaOsci} 
\end{align}
where we inserted the definitions for the tight binding coefficients, shown below Eq. \eqref{eq:piband}. Similar to the work of Sayers, in Eq. \eqref{eq:anaOsci}, bracket right hand side, we clearly find an oscillating behavior in $\mathbf{k}_{\parallel}$ with the lattice vector frequency $\boldsymbol{\delta}_i$ comparable to Eq. \eqref{eq:scat}. Due to the small nuclear number, the broadening of the form factor $f_n$ is that strong that $f_n$ is almost constant, $f_n(\mathbf{k})\equiv f_n$. Therefore, the reciprocal lattice geometry is not resolved and its influence on $\alpha_{\qp}(\omega)$ is negligible. Consequently, Eq. \eqref{eq:anaOsci} is determined solely by the squared tight binding coefficients $(1+\sum_{i}\cos(\mathbf{k}_{\parallel}\cdot\boldsymbol{\delta}_i)/|\xi_{\mathbf{k}_{\parallel}}|)$, which are responsible for the cosine-like oscillations in Fig. \ref{fig:exafs}. From a physical point of view, the tight binding coefficients describe the orbital contribution of a specific lattice site to the band composition, cf. Eq. \eqref{eq:TightBind}. Due to the lattice periodicity, the tight binding coefficients are a function of the wave vector $\kp$, which is determined by the lattice geometry. We can therefore understand the absorption, summing over all wave vectors (Eq. \eqref{eq:susc}) and proportional to Eq. \eqref{eq:anaOsci}, as showing the quantum interference between orbital electronic wave functions, which can be constructive or destructive depending on the sign of the tight binding coefficients. 

Summarizing this discussion, we can conclude that the EXAFS oscillations stem from an interference of standing electronic wave functions of the graphene sublattices $A$ and $B$, which are summed up in Eq. \eqref{eq:anaOsci}. \textit{The interference is a spatially stationary property of the intrinsic electronic wave function, and therefore it is independent of the X-ray excitation}. 

\begin{figure}[t]
\begin{center}
\includegraphics[width=\linewidth]{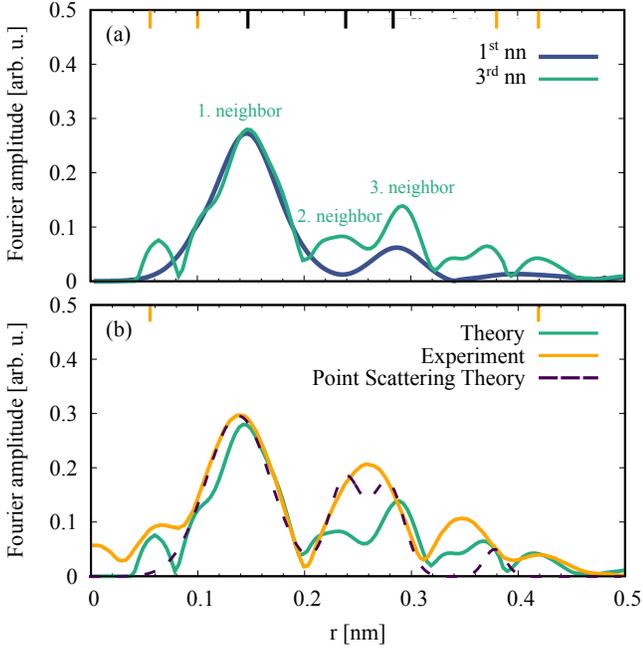}
\end{center}
\caption{(a) Fourier amplitude of the EXAFS spectrum. The blue curve includes only the first nearest-neighbor (nn). It peaks at the first nearest-neighbor distance and its higher harmonics. The green curve includes also the second and third next-neighbors. The peaks stem from the first three neighbors (black) and interferences among themselves (yellow). (b) Comparison of Bloch theory, experiment (adapted from Ref. \cite{Buades2018}) and point scattering theory (Eq. \eqref{eq:FTpointscat}).}
\label{fig:fourier}
\end{figure} 
Since for EXAFS, the X-ray light has an excitation energy higher than the threshold energy, the absorption coefficient $\alpha_{\qp}(\omega)$ can be expressed as function of the wave number of free electrons, i.e. with dispersion $\omega(k)=\hbar k^2/2m_0+E_{Ion}/\hbar$ as final states. To discuss the spatial interference in close analogy to experiments, the absorption coefficient is Fourier transformed $\int d^2k ~ k^m\alpha_{\qp(\kp)}(\kp)\exp(i\kp\cdot\mathbf{r}_{\parallel})$ to real space with respect to the wave number. To increase the visibility and access more conveniently the oscillations at higher $k$, the absorption spectrum as function of wave number is multiplied by $k^m$ with typically $m\in [1,2,3]$ and Fourier transform subsequently \cite{Ertl2008,Kas2016,Britz2020}. In the following we discuss the Fourier transform of the absorption spectrum evaluated by our description, which involves the Bloch theorem for solid states:

The blue curve in Fig. \ref{fig:fourier}(a) displays the Fourier amplitude of the $k$ ($m=1$) weighted and background corrected  EXAFS spectrum \cite{Ravel2005} in first neighbor approximation. The effective radial distribution function clearly displays a peak at the distance of the 1$^{\text{st}}$ neighbor. The second peak at \unit[0.28]{nm} lying at twice of the 1$^{\text{st}}$ nearest-neighbor distance originates from the performed discrete Fourier transformation on a finite grid.

So far we have considered only the nearest-neighbor hoppings for the calculation of the matrix elements. As a result we obtain a peak at the first neighbor distance. However, calculating the matrix elements beyond the nearest-neighbor approximation should add additional features to the Fourier amplitude of the EXAFS, cf. Fig. \ref{fig:fourier}(a) green curve. To go beyond the nearest-neighbor hopping we introduce the second $\xi^{(2)}_{\kp}=\eta_2\sum_{i=4}^{9}\exp(i\kp\cdot\boldsymbol{\delta}_i)$ and third $\xi^{(3)}_{\kp}=\eta_3\sum_{i=10}^{12}\exp(i\kp\cdot\boldsymbol{\delta}_i)$ next-neighbor form factors. For the free parameters of second and third next-neighbor hoppings we reasonably choose $\eta_2=0.05$ and $\eta_3=0.01$, respectively: The ratios of $\eta_2$ and $\eta_3$ with respect to the hopping parameter to the nearest-neighbor are chosen such that they coincide with the relative deviation of the hoppings between 2$p_z$-orbitals with increasing order of neighbors \cite{Kundu2017,Tran2017}. The peak heights are uncertain under this assumption, but the peak positions -- which is the most important in our study -- are unaffected by that. The tight binding coefficients up to the third neighbor hopping read
\begin{align}
C^{1s}_{A1s,\kp}=\frac{\xi_{\kp}+\xi^{(3)}_{\kp}}{\sqrt{2}\left(|\xi_{\kp}+\xi^{(3)}_{\kp}|+\xi^{(2)}_{\kp}\right)}, \quad\text{and}\quad C^{1s}_{B1s}=\frac{1}{\sqrt{2}} \label{eq:TB_3NN}
\end{align}
and enter the matrix element, described by the first line of Eq. \eqref{eq:anaOsci}, now including 12 neighbors instead of three.

The green line in Fig. \ref{fig:fourier}(a) displays the effective radial distribution function including hoppings up to the third neighbor coupling in the EXAFS matrix element. In contrast to the calculation with first neighbor hopping (blue curve), we find additional peaks at the second and third neighbor distance of \unit[0.24]{nm} and \unit[0.28]{nm}, respectively. Interestingly, since the modulus square of the dipole matrix element is observed, also peaks appear, which do not correspond to next neighbor vectors, but do correspond to interferences of different next neighbor vectors. Exemplary, we obtain a peak at \unit[0.38]{nm} corresponding to the sum of the first and second neighbor distance. Since \unit[0.38]{nm} also correspond to the $4^{\text{th}}$ neighbor hopping in graphene, this peak could also be interpreted as the fourth neighbor in experiments. \textit{At \unit[0.05]{nm} we can resolve a peak, which arises from the difference of third and second neighbor. So far, this peak has been explained as phase shift stemming from a difference between measured and geometrical interatomic distances and required a theoretical or experimental correction \cite{Rehr2000}. In contrast, in our approach, we show that the peak can be interpreted as quantum interference between electronic Bloch wave functions of first and second neighbor.} The slight sideband of the first nearest-neighbor peak at \unit[0.1]{nm} stems from the difference of the second and first neighbor. Further, we observe a peak at \unit[0.42]{nm}, which can be understood as interference of the $1^{\text{st}}$ with $3^{\text{rd}}$ neighbor.

To manifest our interpretation, in Fig. \ref{fig:fourier}(b) we provide a direct comparison of our full computational result up to the third nearest-neighbor to experiment, adapted from Ref. \cite{Buades2018}, and additionally display the outcome obtained from the point scattering theory \cite{Sayers1971}. In point scattering theory the oscillatory part of the EXAFS is solely described by the matrix element squared corresponding to Eq. \eqref{eq:scat}. The Fourier transform yields the structure-related function \cite{Sayers1971}
\begin{align}
S(r)=\frac{1}{2}\sum_i \frac{N_i}{\sigma_iR_i^2} e^{-\gamma R_i} e^{-2(r-R_i)^2/\sigma_i^2} \label{eq:FTpointscat}
\end{align}
displayed as dashed line in Fig. \ref{fig:fourier}(b). The $\gamma$-factor accounts for the photoelectron scattering range and leads to an overall decrease with increasing distance. The result is a sum of Gaussian functions lying at the next-neighbor distances $R_i$.

The blue curve in Fig. \ref{fig:fourier}(b) corresponds to the Fourier transform of a measured EXAFS with an isolated soft X-ray pulse produced by high harmonic generation. In the experimental curve, the Fourier transformation the EXAFS data were background corrected. This correction consists of an approximation of the EXAFS data by an adjustable smooth function, which represents the absorption coefficient without neighboring atoms. The spline function is then subtracted from the measured data. Details of the experiment are given in the SM Sec. X. To have a fair comparison between the two theories (point scattering and Bloch theorem based approach) and the experiment we perform the same manipulations. Additionally, we force the first neighbor peak to match the experimental one in height. The experimental curve in Fig. \ref{fig:fourier}(b) displays three major peaks: The first at \unit[0.14]{nm} reflects the 1$^{\text{st}}$ neighbor. The second peak around \unit[0.26]{nm} consists of a sum of 2$^{\text{nd}}$ and 3$^{\text{rd}}$ neighbor. The reason that in the experiment the 2$^{\text{nd}}$ and 3$^{\text{rd}}$ neighbor peak add to one spectral signature is that the momentum transfer range of the measurement is limited, which smears out the Fourier transformed peaks. We expect that this effect could be minimized with a broader spectral range of the laser. Lastly, the maximum at \unit[0.38]{nm} can be interpreted as 4$^{\text{th}}$ neighbor. All three peaks are well reproduced by our Bloch as well as the point scattering theory. However, a close look to the experiment reveals clear additional spectral features around \unit[0.05]{nm} and at \unit[0.42]{nm}, both distances that do not exist in the graphene lattice. While those spectral features are absent in a point scattering theory, our theory reproduces them in good agreement in position: Following the description above, these features rely on the use of the Bloch theorem and can be explained as quantum interference between electronic Bloch wave functions. Although the interference peaks are observed in experiment \cite{Buades2018} they have not been discussed so far. Moreover, the existing interpretation in literature is not able to explain these spectral features, but a detailed solid state theory involving the full solid state lattice symmetry is required to interpret these features. Thus a first central result of our approach is a solid state generalization of Eq. \eqref{eq:scat}.

Summarizing, the EXAFS oscillations are encoded in the dipole matrix element, which modulate the square root-like absorption line into the three-dimensional continuum. The modulations can only be understood by a solid state specific theory that confirms the relation between the EXAFS oscillations and the local real space configuration of the crystalline material \cite{Sayers1971}. However, these oscillations does not originate from an interference of the X-ray waves but from quantum interferences of electronic wave functions of neighbored atoms.

\begin{figure}[t]
\begin{center}
\includegraphics[width=\linewidth]{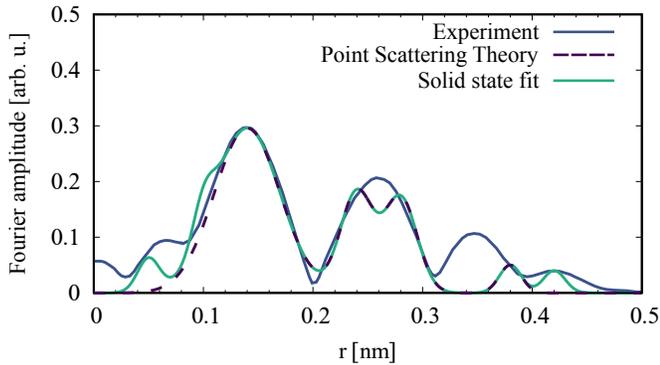}
\end{center}
\caption{Comparison with the newly suggested fit formula. The suggested expression for the structure-related function includes now also the phenomenological phase shift peak and the peak at \unit[0.42]{nm} in addition to the true next-neighbor distances. } 
\label{fig:fourier2}
\end{figure}

With the newly obtained insights, we want to finally suggest a phenomenological modified fit formula for EXAFS of crystalline solids. As we identified interference peaks, we suggest to add to Eq. \eqref{eq:scat}:
\begin{align}
\alpha_{k}&=S_0^2\sum_i \frac{N_i|f_i|}{k R_i^2} e^{-\gamma_iR_i} [e^{-2\sigma_i^2k^2}  \sin\left(2kR_i\right)  \nonumber \\
&+S_0^2\sum_{j>i,\pm} \frac{N_j|f_j|}{kR_j^2} e^{-\gamma_j R_j}e^{-2\sigma_j^2k^2} \sin(2k|R_i \pm R_j|) ] \label{eq:Sayer_new}
\end{align}
where we abbreviated the finite lifetime of the photoelectron by $\gamma$. In the newly added term, the $j$ sum runs over all next neighbor orders higher than $i$. In the SM Sec. XI, we calculate the absolute square of the sum of the tight binding coefficients Eq. \eqref{eq:TB_3NN} which support the inclusion of the interference peaks in a form of the second line of Eq. \eqref{eq:Sayer_new}. A Fourier transformation yields for the structure-related function
\begin{align}
S(r) &= \sum_i \frac{N_i}{\sigma_i R_i^2} e^{-\gamma_iR_i} [ e^{-2(r-R_i)^2/\sigma_i^2} \nonumber \\
&+ \sum_{j>i,\pm}  \frac{N_j}{\sigma_jR_j^2} e^{-\gamma_j R_j} e^{-2(r-|R_i\pm R_j|)^2/\sigma_j^2} ]  \label{eq:FT_new}
\end{align}
where the second summand now attributes for the interference peaks. Figure \ref{fig:fourier2} compares the original solution Eq. \eqref{eq:FTpointscat} with the experiment and the solid state specific suggestion Eq. \eqref{eq:FT_new}. For both fit functions we used the same parameters as for Fig. \ref{fig:fourier}. We see that we can now address also the phase shift peak rising before the first neighbor and the peak at \unit[0.42]{nm}.
 
\section{Conclusion}

We have presented a self-consistent theory of Maxwell and Bloch equations to describe X-ray absorption in crystalline two-dimensional crystals by incorporating the Bloch theorem and many-body effects. The occurring description develops XANES, EXAFS, excitonic, core-hole, and nonlinear effects as well as radiative and Meitner-Auger recombination out of one Hamiltonian. As an example, we study the linear response, where in-plane excited Bloch waves interfere in the total susceptibility. Also, due to the interplay of the polarization of the underlying electron transitions, we find a significant dependence on the angle of the incident light, which differs for different spectral regions. By this, we developed a XANES description, which goes beyond the usually used Fermi's golden rule. We demonstrated that the XANES part of the spectrum maps the density of states of bound electronic states, whereas in the EXAFS substantial oscillations can be found, which can be identified as transitions between different atoms in the layer. Consequently, the Fourier transformed EXAFS spectrum exhibits peaks at the distances between atoms of consecutive atomic sites and at sums or differences of them. This corresponds to first microscopic insights to the origin of these spectral oscillations in two-dimensional solid states. Further, our approach assign so far overlooked peaks in the Fourier transformed EXAFS spectrum.

The line broadening occuring in optical spectra is related to the core-hole lifetime. Two prominent recombination mechanisms are radiative and Meitner-Auger recombination. The radiative recombination can be described in a self-consistent way by combining the wave equation and the microscopic Bloch equation under consideration of the in-plane X-ray wave vector. The Meitner-Auger contributions result straightforward from the Coulomb interactions within the Bloch equation formalism. With this, we give an extension of the typical core-hole recombination channels in atomic systems to solid states. The presented X-ray absorption theory for two-dimensional materials can be straightforwardly expanded to layered materials, by adjusting the definition of the two-dimensional polarization by adding a layer index. The strength of the developed method relies on its possibility to be combined with \textit{ab initio} electronic structure theory (DFT) to explicitly calculate all input matrix elements.

So far for this contribution, we focused on X-ray absorption spectroscopy. An additional quantization of the electric field yields a fully quantized light-matter Hamiltonian, which enables the description of X-ray fluorescence. Further so far, the absorption coefficient depends on the X-ray wave vector $\qp$, which equals the incident X-ray wave vector. However, investigating the absorption coefficient as a function of a different X-ray wave vector $\qp'$ leaving the sample, would lead to the possibility of investigating the absorption cross section additionally as function of the scattering angle as done for example in resonant elastic or inelastic X-ray scattering (REXS or RIXS) \cite{Wang2020,Monney2020,Vorwerk2022}. However, as pointed out for XANES this requires a close connection with DFT methods for single-particle energies to accurately describe the experiments.

\vspace{5mm}
We thank Florian Katsch and Robert Salzwedel (TU Berlin) for fruitful discussions. We acknowledge financial support from the Deutsche Forschungsgemeinschaft (DFG) through SFB 951 (A.K., D.C., M.S.) Projektnummer 182087777. This project has also received funding from the European Unions Horizon 2020 research and innovation programme under Grant Agreement No. 734690 (SONAR, A.K.). D.C. thanks the graduate school Advanced Materials (SFB 951) for support. J.B. acknowledges financial support from the European Research Council for ERC Advanced Grant “TRANSFORMER” (788218), ERC Proof of Concept Grant “miniX” (840010), FET-OPEN “PETACom” (829153), FET-OPEN “OPTOlogic” (899794), Laserlab-Europe (654148), Marie Sklodowska-Curie ITN “smart-X” (860553), MINECO for Plan Nacional FIS2017-89536-P; AGAUR for 2017 SGR 1639, MINECO for “Severo Ochoa” (SEV- 2015-0522), Fundació Cellex Barcelona, the CERCA Programme / Generalitat de Catalunya, and the Alexander von Humboldt Foundation for the Friedrich Wilhelm Bessel Prize.

\bibliographystyle{ieeetr}


\end{document}